\documentclass[11pt,a4paper]{emulateapj}

\newcommand{\Mpc}{~h^{-1}~\rm{Mpc}}
\newcommand{\Msun}{~h^{-1}~M_{\odot}}
\newcommand{\den}{~h^{3}~\rm{Mpc^{-3}}}
\newcommand{\gps}{\ensuremath{g_{\rm P1}}}
\newcommand{\rps}{\ensuremath{r_{\rm P1}}}
\newcommand{\ips}{\ensuremath{i_{\rm P1}}}
\newcommand{\zps}{\ensuremath{z_{\rm P1}}}
\newcommand{\yps}{\ensuremath{y_{\rm P1}}}

\newcommand{\grizy}{\gps\rps\ips\zps\yps}

\slugcomment{}

\shorttitle{PFOF}

\shortauthors{Jian et al.}

\begin{document}

\title{Probability Friends-of-Friends (PFOF) Group Finder: Performance Study and Observational Data Applications on Photometric Surveys}

\author{Hung-Yu Jian\altaffilmark{1}$^{\star}$, Lihwai Lin\altaffilmark{2}, Tzihong Chiueh\altaffilmark{1,9,10}, Kai-Yang Lin\altaffilmark{2}, Hauyu Baobab Liu\altaffilmark{2}, Alex Merson\altaffilmark{3,4}, Carlton Baugh\altaffilmark{4}, Jia-Sheng Huang\altaffilmark{5,11}, Chin-Wei Chen\altaffilmark{2}, Sebastien Foucaud\altaffilmark{6}, David N. A. Murphy\altaffilmark{7}, Shaun Cole\altaffilmark{4}, William Burgett\altaffilmark{8}, and Nick Kaiser\altaffilmark{8}}
\email{$\star$ E-mail: hyj@phys.ntu.edu.tw}

\altaffiltext{1}{Department of Physics, National Taiwan University, 106, Taipei, Taiwan, R.O.C.}
\altaffiltext{2}{Institute of Astronomy \& Astrophysics, Academia Sinica, 106, Taipei, Taiwan, R.O.C.}
\altaffiltext{3}{Department of Physics and Astronomy, University College London, Gower Street, London WC1E 6BT, UK}
\altaffiltext{4}{Institute for Computational Cosmology, Department of Physics, Durham University, South Road, Durham DH1 3LE, UK}
\altaffiltext{5}{National Astronomical Observatories, Chinese Academy of Sciences, Beijing 100012, China}
\altaffiltext{6}{Department of Earth Sciences, National Taiwan Normal University, N.88, Tingzhou Road, Sec. 4, Taipei 11677, Taiwan, Republic of China}
\altaffiltext{7}{Department of Astronomy and Astrophysics, Pontificia Universidad Cat\'{o}lica de Chile, Vicuna Macken$\widetilde{n}$a 4860, 7820436 Macul, Santiago, Chile
}
\altaffiltext{8}{Institute for Astronomy, University of Hawaii, 2680 Woodlawn Drive, Honolulu HI 96822}

\altaffiltext{9}{Center for Theoretical Sciences, National Taiwan University, 106, Taipei, Taiwan, R.O.C. }
\altaffiltext{10}{LeCosPa, National Taiwan University, 106, Taipei, Taiwan, R.O.C.}
\altaffiltext{11}{Harvard-Smithsonian Center for Astrophysics, 60 Garden Street, Cambridge,
MA 02138, USA}

\begin{abstract}
 In tandem with observational datasets, we utilize realistic mock catalogs, based on a semi-analytic galaxy formation model, constructed specifically for Pan-STARRS1 Medium Deep Surveys in order to assess the performance of the Probability Friends-of-Friends (PFOF, Liu et al.) group finder, and aim to develop a grouping optimization method applicable to surveys like Pan-STARRS1. Producing mock PFOF group catalogs under a variety of photometric redshift accuracies ($\sigma_{\Delta z/(1+z_s)}$), we find that catalog purities and completenesses from ``good'' ($\sigma_{\Delta z/(1+z_s)}$ $\sim$ 0.01) to ``poor'' ($\sigma_{\Delta z/(1+z_s)}$ $\sim$ 0.07) photo-$z$s gradually degrade respectively from 77$\%$ and 70$\%$ to 52$\%$ and 47$\%$. To avoid model dependency of the mock for use on observational data we apply a ``subset optimization'' approach, using spectroscopic-redshift group data from the target field to train the group finder for application to that field, as an alternative method for the grouping optimization. We demonstrate this approach using these spectroscopically identified groups as the training set, i.e. zCOSMOS groups for PFOF searches within PS1 Medium Deep Field04 (PS1MD04) and DEEP2 EGS groups for searches in PS1MD07. We ultimately apply PFOF to four datasets spanning the photo-$z$ uncertainty range from 0.01 to 0.06 in order to quantify the dependence of group recovery performance on photo-$z$ accuracy. We find purities and completenesses calculated from observational datasets broadly agree with their mock analogues. Further tests of the PFOF algorithm are performed via matches to X-ray clusters identified within the PS1MD04 and COSMOS footprints. Across over a decade in group mass, we find PFOF groups match $\sim$85$\%$ of X-ray clusters in COSMOS and PS1MD04, but at a lower statistical significance in the latter. This high matched fraction lends additional support to the reliability of the detection algorithm. In the end, we demonstrate, by separating red and blue group galaxies in the EGS and PS1MD07 group catalogs, that the algorithm is not biased with respect to specifically recovering galaxies by color. The analyses in our study suggest the PFOF algorithm shows great promise as a reliable group finder for photometric galaxy surveys of varying depth and coverage.

\end{abstract}


\section{Introduction}
In the $\Lambda$ cold dark matter ($\Lambda CDM$) cosmology, structures grow hierarchically, with small objets forming initially from very small density fluctuations and merging successively to become the large-scale structures we detect today \citep{pee82,blu84,dav85,spr05}. The study of the formation and evolution of large-scale structures is, therefore, important for improving our understanding of the Universe on both cosmological and galactic scales. As the largest bound objects in the Universe, clusters of galaxies are naturally the structures appropriate for such a study. The high mass end of the mass distribution of collapsed structures, i.e. galaxy clusters and groups, is the most accessible observationally, and hence cluster masses often serve as probes to constrain the cosmological parameters \citep[e.g.,][]{car96,bor01,sch03,man08,all11}. Additionally, galaxy clusters are also ideal sites for gravitational lensing studies of the dark matter \citep[e.g.,][]{she04,wit06,sma07} due to the dominance of dark matter over baryons, contributing $\sim$ 85$\%$ of the whole mass content. Clusters not only have strong X-ray signatures due to the hot intracluster gas trapped inside deep gravitational potentials \citep{sar86}, but the hot gas also leaves an imprint on the cosmic microwave background (CMB) through the Sunyaev-Zel'dovich effect \citep{sun80,car02,lin04a}. Many studies indicate that clusters are important laboratories for investigating galaxy evolution \citep[e.g.,][]{oem74,lin04,voi05}.

Moreover, some studies have suggested the presence of a critical halo mass, roughly corresponding to the mass of the group scale, above which star formation is efficiently quenched \citep[e.g.,][]{dek06,gil08}. The group environment is also an important location for studying galaxy formation and evolution. Observationally, it has been found that the galaxy merger rate depends strongly on environment \citep{lin10,deRav11,kam13}, and evidence from semi-analytical models implemented in the Millennium Simulation shows that galaxies in denser environments tend to have higher merger rates than those in the fields, and these higher density regions are primarily dominated by the group environments \citep{jian12}. It is therefore of great importance to produce reliable catalogs of galaxy groups and clusters that can be used to study the role of environment in galaxy evolution.

In the optical regime, group identification has been attempted over a number of years. Methods for group finding in galaxy surveys can be roughly divided into two categories: methods using spectroscopic data and those using photometric data. Spectroscopic galaxy redshift surveys can reduce much of the projection effect problem, except for the fingers-of-God effect due to galaxy motion inside virialized structures. The Friends-of-friends (FoF) algorithm \citep{huc82} is one of the most popular approaches to group finding and is still in common use in present-day redshift surveys \citep{kno09,kno12}. FOF plus simple assumptions about the properties of galaxies in groups and clusters is also in use \citep[e.g.,][]{eke04,yan05}. In addition, the Voronoi-Delaunay Method (VDM) of \cite{mar02}, which is claimed to compensate for some of the shortcomings of the traditional FOF algorithm, is also a common method used for group identification. This latter method was adopted by the DEEP2 collaboration \citep{ger05,ger12}. The sparse sampling of galaxies in spectroscopic surveys is one known shortcoming of group-finding, and is a major concern for high redshift surveys, which are observationally expensive.

By contrast, photometric surveys have a much more complete sampling rate, but the foreground and background contamination can be significant especially for modern deep photometric surveys. Therefore, group finding algorithms for this type of survey typically must include additional components to alleviate such contaminations. For example, algorithms can rely on assumptions about the properties of galaxies in clusters, such as the matched-filter or red-sequence methods \citep[e.g.,][]{pos96,gla00,koe07,mur12}. Algorithms can take into account photometric redshift estimates \citep[e.g.,][]{li08,gil11}, including the Probability Friends-of-Friends (PFOF) algorithm \citep{liu08}\footnote{\cite{li08} also developed a group searching method in photometric-redshift space with the identical name, the probability Friends-of-Friends algorithm, to ours. However, two grouping methods have distinct algorithms and selection functions. We stress here that the reader should be cautious to get confused by the name.}. Algorithms can also be based on a combination of the two \citep[e.g.,][]{mil10}. Red-sequence methods \citep{gla00,koe07,mur12} utilize the presence of the red-sequence ridgeline, and have been demonstrated to be successful at detecting clusters. However, their advantages can also turn into drawbacks. The identification of structures tends to be biased toward rich clusters with red galaxies. In addition, at high redshift, the presence of the red-sequence ridgeline is less clear, making red-sequence methods less reliable. Compared to red-sequence methods, PFOF only needs information regarding the location of galaxies and photometric redshift estimates without preferential color selection, and can on one hand include blue members which are missed by red-sequence methods, whilst on the other hand identify groups of low richness with no red-sequence galaxies. Therefore, PFOF has its own unique advantages.

In the past few decades, large area, deep optical and infrared galaxy surveys have shown rich group and cluster structures on various scales. Nowadays, even greater area and deeper multiband optical sky surveys, such as Pan-STARRS \citep{kai02} or the Dark Energy Survey \citep{fri13}, have been carried out to image more of the Universe. Forthcoming surveys such as those with the Subaru Hyper Suprime-Cam \citep{tak10} or the Large Synoptic Survey Telescope \citep{ive08} will also start soon. With such a huge amount of data, a reliable group finder is obviously needed for group and cluster studies. We shall show in this paper that PFOF is a highly competitive tool for this purpose.

This work is the extension of \cite{liu08} (hereafter Liu08). In Liu08, the algorithm, Probability Friends-of-Friends (PFOF), was presented and simple tests based on DEEP2 mock catalogs were carried out. In this paper, we extend Liu08 by adding further tests of the performance of PFOF. We adopt an improved Durham Pan-STARRS1 mock catalog, which has a larger field of view and greater depth, for use in various tests to demonstrate the performance of PFOF. In additional to scaling the linking length to the mean separation of galaxies as a function of redshift, we also adopt an optimization measure introduced by \cite{kno12} to produce the optimal catalogs. Moreover, we apply ``subset optimization'' which utilizes spectral-$z$ groups as a training set, to the observational data to optimize the PFOF group catalog, so as to avoid any model dependency through using the mock. Datasets with ``good'' photometric redshift accuracy, $\sigma_{\Delta z/(1+z_s)}$ $\sim$ 0.01 in COSMOS, ``medium'' accuracy  $\sigma_{\Delta z/(1+z_s)}$ $\sim$ 0.03 in EGS, and relatively ``poor'' accuracy $\sigma_{\Delta z/(1+z_s)}$ $\sim$ 0.06 in Pan-STARRS1 Medium-Deep04 (PS1MD04) and PS1MD07 are tested to understand the dependence of the optimal performance on photo-$z$ accuracy and to evaluate the applicability of PFOF. Finally, results are also compared with an X-ray catalogs to demonstrate the success of the subset optimization and the PFOF capability.

This paper represents the first in a series of studies using PS1 medium deep surveys (MDS) based on the version of data reduction provided by Foucaud et al. (2013, in preparation). In Lin et al. (2013, in preparation), the environmental effect on the star formation rate in PS1 MDS data will be studied. This paper is structured as follows. In Section 2, we briefly describe the datasets we use, including the Durham mock, one X-ray, two spectral-$z$ catalogs, and four photometric-redshift galaxy catalogs. The PFOF algorithm as well as definitions of purity and completeness are then introduced in Section 3. Tests of the mock data for PFOF are illustrated in Section 4. In Section 5, observational data applications and performance tests for PFOF are provided. Finally, we give our summary and discussion in Section 6. Throughout this paper, we adopt the following cosmology: $H_0$ = 100 $h\rm{~km~s^{-1}~Mpc^{-1}}$, where the Hubble constant $h$ = 0.73, matter density $\Omega_m$ = 0.24, and cosmological constant $\Omega_{\Lambda}$= 0.76.

\begin{table*}
  \begin{center}
  \leavevmode
  \caption{Datasets}\label{tbl:data}
    \begin{tabular}{llccccll} \hline\hline
    Catalog & Catalog & Area    & Redshift & Flux Limit & $\sigma_{\Delta z/(1+z_s)}$ & Purpose & Ref \\
    Name    & Type    & (deg$^2$) &          &            &                 &         &     \\ \hline
    Durham Mock& Mock      & $\sim$ 7 &  $z$ $<$ 3.0 & $i \leq$ 25.8 & & Performance  & (1) \\
    \cline{1-8}
    COSMOS     & Photometric & $\sim$ 2 & 0 $<$ $z$ $<$ 1.25 & $i^{+}_{AB} \leq$ 24.2 &$\sim$ 0.0125  &Input  & (2) \\
     &  &  &  &  & ($i^{+}_{AB} \leq$ 24) &  \\

    EGS        & Photometric & $<$ 0.5 & 0 $<$ $z$ $<$ 1.4 & $i \leq$ 24.1 & $\sim$ 0.025 to  &Input & (3)  \\

    &  &  &  &  & ($R$ $\leq$ 24.1) &  \\

    PS1MD04    & Photometric & $\sim$ 7 & 0 $<$ $z$ $<$ 1.4 & $\ips \leq$ 24.1 & $\sim$ 0.46  &Input  & (4) \\

    &  &  &  &  & ($i_{\rm P1}$ $\leq$ 22.5) &  \\

    PS1MD07    & Photometric & $\sim$ 7 & 0 $<$ $z$ $<$ 1.4 & $\ips \leq$ 24.1 & $\sim$ 0.51  &Input & (4)  \\

    &  &  &  &  & ($i_{\rm P1}$ $\leq$ 24.1) &  \\
    \cline{1-8}
    zCOSMOS   & Spectroscopic& $\sim$ 1.7 & 0 $<$ $z$ $<$ 1.2 & $i_{AB} \leq$ 22.5 &      & Training  & (5)\\

    DEEP2 EGS   & Spectroscopic& $\sim$ 0.5 & 0 $<$ $z$ $<$ 1.4 & $R_{AB} \leq$ 24.1 &      & Training & (6)\\
    \cline{1-8}
    XMM-NEWTON+Chandra  & X-ray & $\sim$ 2 & 0 $<$ $z$ $<$ 1.0 &  &      & Performance   & (2) \\
   \hline
    \end{tabular}
  \end{center}
  \tablerefs{
(1) \cite{mer13}; (2) \cite{geo11}; (3) Huang et al. (2013); (4) Foucaud et al. (2013, in preparation); (5) \cite{kno09}; (6) \cite{ger12}
}
\end{table*}

\section{Data}
\subsection{Mock Pan-STARRS Medium Deep Survey catalog}
The mock catalog that we use to evaluate the performance of our group finder is based on the Millennium dark matter N-body simulation \citep{spr05}. Halos in the simulation are first identified using a FOF halo finder \citep{dav85} with a linking length of $b$ = 0.2 in units of the mean particle separation. Each FOF-identified halo is then broken into constituent subhalos by the SUBFIND algorithm \citep{spr01}, which identifies gravitationally bound substructures within the host FOF halo. With all halos and subhalos determined, the hierarchical merging trees containing the details of how structures build up over cosmic time can then be constructed. These trees are the key information needed to compute the physical properties of the associated galaxy population for semi-analytical models. The mock catalog adopts the \cite{lag12} model which takes advantage of the extension to the treatment of star formation introduced into GALFORM \citep{col00} in \cite{lag11} to populate galaxies, and is then assembled into a lightcone \citep{mer13}. The dataset covers an area of 50.25 deg$^2$ and includes PS1 $\grizy$ photometry for galaxies down to a magnitude limit of $\ips$ $<$ 25.8 and a redshift range up to $z$ $\sim 3$. A central area of $\sim$ 7 deg$^2$ has been selected for our analyses, which is equivalent to a single pointing of PS1, the area of a single Medium Deep Survey (MDS) tile. This mock field contains 1,601,486 galaxies with $\ips$ $<$ 25.8 and 7,756 groups, i.e. the halos in the simulation, with richness $N$ $\geq$ 4.

The semi-analytic models are tuned to match a selection of observations of local galaxies, e.g. the field galaxy luminosity function, but are not explicitly tuned to match any group data. It is plausible that two models which match the field luminosity function could have different group properties, and may therefore suggest different parameters for the group finder. In this paper, the mock catalog is thus used only to explore the tendency of the grouping performance of PFOF under various photo-$z$ accuracies. For real data application, we will not use the parameters obtained from the mocks.

\subsection{The COSMOS survey}
In the 2-deg$^2$ COSMOS field \citep{sco07}, the galaxy catalog adopted in our analysis is from the X-ray group membership galaxy catalog\footnote{The catalog can be downloaded via the link: http://irsa.ipac.caltech.edu/data/COSMOS/tables/groups/.} described in \cite{geo11}. The photo-$z$ estimation uses an updated version ($\tt pdzBay$\_$\tt v1.7$\_$\tt 010809$) presented in \cite{ilb09} with additional deep $H$-band data and small improvements in the template-fitting techniques. The precise redshifts in the catalog are computed with 30 broad, intermediate, and narrowbands covering the UV (Galaxy Evolution Explorer), visible near-IR (NIR; Subaru, Canada-France-Hawaii Telescope (CFHT), United Kingdom Infrared Telescope, and National Optical Astronomy Observatory), and mid-IR (Spitzer/IRAC). A redshift dispersion of $\sigma_{\Delta z/(1+z_s)}$ is $\sim$ 0.007 at $i^{+}_{AB}$ $<$ 22.5, and at fainter magnitudes $i^{+}_{AB}$ $<$ 24 and $z$ $<$ 1.25, the accuracy $\sigma_{\Delta z/(1+z_s)}$ is $\sim$ 0.012 \citep{ilb09}.  The catalog contains 115,844 galaxies with $i$ $\leq$ 24.2 and 129 X-ray groups with $\tt FLAG$\_$\tt INCLUDE$ = 1, indicating that these groups have high X-ray quality, more than three members, no mask, and no merger signature \citep{geo11}, and is adopted in this paper for performance assessment of surveys with ``good'' photo-$z$ accuracy.

\subsection{The zCOSMOS survey and Group Catalog}
zCOSMOS \citep{lil07} is a spectroscopic redshift survey covering the 1.7-deg$^2$ of the COSMOS field, and consists of two parts, ``zCOSMOS-bright", a pure magnitude selected survey with 15 $\leq$ $I_{AB}$ $\leq$ 22.5, and ``zCOSMOS-deep", aiming at observing about 10,000 galaxies in the redshift range 1.5 $<$ $z$ $<$ 3.0 selected through a well defined color criteria. zCOSMOS-bright, which covers mainly the redshift range 0.1 $<$ $z$ $<$ 1.2, almost the entire COSMOS field, is complete and contains spectra of about 20,000 objects taken using the VIMOS spectrograph with a medium-resolution grism. \cite{kno09} construct an optical group catalog (the 10k catalog) between 0.1 $<$ $z$ $<$ 1 based on $\sim$ 8,417 high-quality spectroscopic redshifts in the zCOSMOS-bright survey. \cite{kno12} recently released an updated optical group catalog (the 20k catalog) covering 0.1 $\leq$ $z$ $\leq$ 1, based on 16,500 high-quality spectroscopic redshifts in the completed zCOSMOS-bright survey. However, the 20k catalog includes only group galaxies without field galaxies so that it can not be adopted for training purposes. We thus make use of this 10k group catalog as the training set to optimize our linking lengths and threshold for datasets covering the COSMOS field. The catalog contains 802 groups with richness $N$ $\geq$ 2.

\subsection{The EGS photometric redshift catalog}
The EGS photometric redshift catalog has a Field-of-View of 0.5-deg$^2$ and is based on the photometric observations in the extended Groth strip (EGS) consisting of 18 bands from $u$ to 8 $\mu$m \citep{hua13}. After combining redundant bands, there are 12 wavelengths available for the photometric redshift estimation. The photo-$z$ accuracy for this catalog is on the order of $\sim$ 0.025 with 3.5$\%$ outliers. The catalog contains 11,229 galaxies, and is used as for the performance test with ``medium'' photo-$z$ accuracy.

\subsection{The DEEP2 survey and Group Catalog}
The DEEP2 Galaxy Redshift Survey \citep{dav03,new12} is a spectroscopic survey covering a combined area of four separate fields of approximately 3-deg$^2$ down to a limiting magnitude of $R_{AB}$ $<$ 24.1, and probes a volume of 5.6 $\times$10$^6$ $\den$ over the primary DEEP2 redshift range 0.75 $<$ $z$ $<$ 1.4. There are 50,000 spectra obtained in 1 hr exposures with the DEIMOS spectrograph \citep{fab03} on the Keck II telescope, and in this dataset 35,000 objects are confirmed with galaxy redshifts. Overall, the sampling rate is roughly 70$\%$ of the median sampling rate in the most crowded regions, and the redshift success rate is also about 70$\%$ \citep{new12}.

\cite{ger12} present a public catalog of galaxy groups constructed from the spectroscopic sample of galaxies using Voronoi-Delaunay method (VDM) in the fourth data released from DEEP2 Galaxy Redshift Survey, including EGS. In the EGS field, the catalog contains 12,346 galaxies in the redshift range z = 0 to 1.4 down to  $R_{AB}$ $\leq$ 24.1 and 1,165 groups with richness $N$ $\geq$ 2. The EGS field overlaps with the Pan-STARRS 1 Medium Deep Field 07 (PS1MD07) and is used as a training set for the optimization of PFOF.

\subsection{The Pan-STARRS1 Medium Deep Survey}
Pan-STARRS1 \citep{kai02,kai10,cha11} is a 1.8 m optical telescope with a 7 square degree field of view that can image the sky in $\gps$, $\rps$, $\ips$, $\zps$, and $\yps$ filters which cover the 4000{\AA} $<$ $\lambda$ $<$ 10500{\AA} spectral range \citep{stu10,ton12}. Images obtained by the Pan-STARRS1 system are processed through the Image Processing Pipeline \citep[IPP;][]{mag06}. Photometric and astrometric measurements performed by the IPP system are described in \cite{mag07} and \cite{mag08}, respectively.

The official deep stacks for the Medium Deep fields from IPP were not available for use. We instead adopt Foucaud's deep stacks (Foucaud et al. 2013, in preparation), which include the $\grizy$ bands for all nightly stacks from April 2010 to December 2011, plus the CFHT $u$-band from archival data. Astrometric and photometric calibration are performed with SDSS-DR7, using SCAMP \citep{ber02}. A 4 sigma-clipped median stacking is performed with SWarp \citep{ber02}. The photometric catalog is then extracted with SExtractor \citep{ber96}, in dual-mode with i-band as the detection image. Photo-$z$ are computed with EASY \citep{bra08}, using a prior on the redshift distribution at a given i-band magnitude from the semi-analytical model of \cite{guo11} and with zero-point (ZP) corrections applied (Lin et al. 2013, in preparation). Two fields MD04 (or PS1MD04) and MD07 (or PS1MD07) are selected as examples for this first study. PS1MD04 has a photo-$z$ accuracy $\sim$ 0.046 with a outlier rate 0.038 down to $i_{\rm P1}$ $<$ 22.5. PS1MD07 has a photo-$z$ accuracy $\sim$ 0.051 with a outlier rate 0.071 down to $r_{\rm P1}$ $<$ 24.1. The outlier rate is defined as the fraction of objects for which $\mid$$z_{phot}$ - $z_{s}$$\mid$ $>$ 0.15 $\times$ (1 + $z_{s}$), and the photo-$z$ accuracy, $\sigma_{\Delta z/(1+z_s)}$, is estimated from the deviation of $\mid$$z_{phot}$ - $z_{s}$$\mid$ / $(1+z_s)$ without the outliers. Two PS1 datasets are adopted to demonstrate the performance of PFOF with relatively ``poor'' photo-$z$ accuracy.

Information of all datasets is summarized in Table~\ref{tbl:data}.

\section{Method}

\subsection{Probability Friends-of-Friends}
\begin{figure}
 \begin{center}
  \includegraphics[angle=0,scale=0.3]{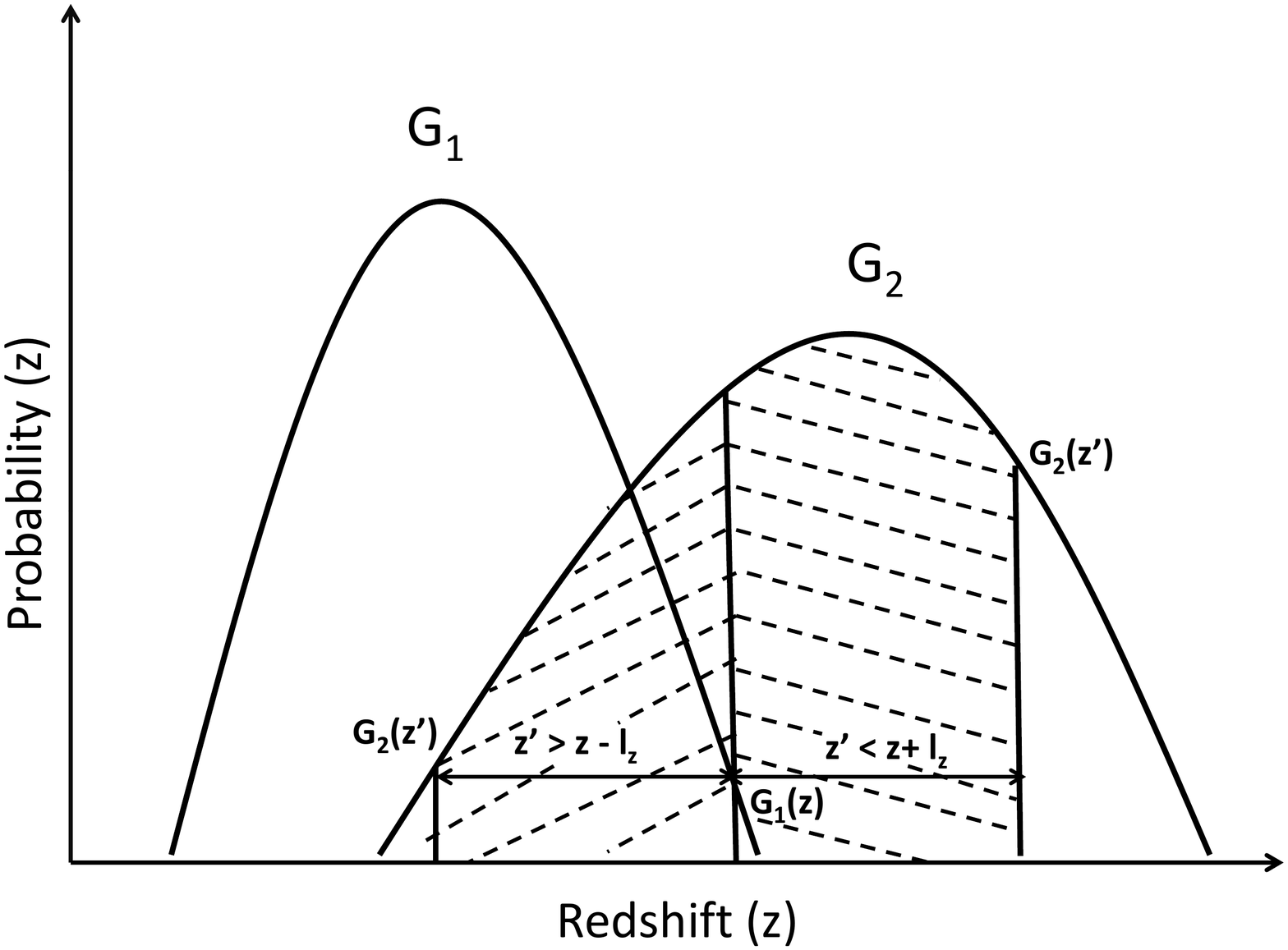}
  \caption{The schematic illustration of Equation~(\ref{prob}), which expresses the probability that how galaxies with photometric redshifts, have a probability of being separated by $|z_2 - z_1|$ which is less than the radial linking length $l_z$.
   }\label{fig:linkprob}
 \end{center}
\end{figure}

Probability Friends-of-Friends (PFOF) was developed by Liu08 to identify galaxy groups and clusters in a galaxy catalog with redshift measurement errors. We briefly review the algorithm below, and readers are referred to Liu08 for a detailed discussion. PFOF is based on the FOF algorithm \citep{huc82}, modified to take into account photometric redshift uncertainty. Therefore, similar to the FOF algorithm, the criteria applied in PFOF to determine if two galaxies are physically linked are divided into two parts, the condition in the projected plane and in the line-of-sight direction. In the projected plane, the linking criterion is to examine whether the separation of two galaxies, $d_{12}$, is less than the comoving linking length $l_{p}$, i.e. $d_{12} \leq l_{p}$, where $l_p$ is a parameter. In the line-of-sight direction, given the photometric redshift probability distribution functions for the two galaxies, $G_{1}$ and $G_{2}$, the probability $P$ of the distance between them being less than the $z$-linking length, $l_z$, is defined as

\begin{equation}\label{prob}
    P(|z_{2} - z_{1}| \leq l_{z}) \equiv \int^{\infty}_{0} dz G_{1}(z) \int^{z+l_{z}}_{z-l_{z}} G_{2}(z^{'}) dz^{'},
\end{equation}
where $l_{z}$ is the comoving linking length in the line-of-sight direction. The linking criterion then has to satisfy
\begin{equation}\label{criterion2}
    P(|z_{2} - z_{1}| \leq l_{z}) \geq P_{th},
\end{equation}
 where $P_{th}$ is a tunable linking probability threshold. When both criteria are satisfied, two galaxies are called friends. The integration is described schematically in Figure~\ref{fig:linkprob}. We continue searching all the friends of one galaxy and then the friends of friends, until finally a group is formed. In this way, given $l_{p}$, $l_{z}$, and $p_{th}$, PFOF constructs a list of group members. For a sample with a limiting magnitude, the mean density of galaxies $n$ decreases with increasing redshifts, leading to a steady increase in the mean inter-galaxy separation with $z$. To compensate for this effect, both $l_p$ and $l_z$ in terms of $n_0^{-1/3}$, where $n_0$ is the mean galaxy density, i.e.
\begin{equation}\label{lp}
 l_p(z) = b_{p} \ n_0(z)^{-1/3},
\end{equation}
 and
 \begin{equation}\label{lz}
 l_z(z) = b_{z} \ n_0(z)^{-1/3},
\end{equation}
where $b_{p}$ and $b_{z}$ are dimensionless linking parameters perpendicular and parallel to the line of sight. Thus, the three adjustable parameters in the PFOF group finder are $b_{p}$, $b_{z}$, $P_{th}$.

\subsection{Purity and Completeness}
To assess the performance of a group finder, two quantities, purity and completeness, can be adopted to characterize the overall fidelity of the resulting group catalogs, i.e. \cite{eke04}. However, different definitions for purity and completeness can lead to distinct results. Our definitions for purity and completeness follow those of \cite{ger05}. In the following, we briefly explain the definitions used. For a more detailed discussion the reader is referred to \cite{ger05}, \cite{kno09}, and \cite{kno12}. We first define that a ``reconstructed group" (or a PFOF group) identified by our group-finder is a ``pure group'', when more than $50\%$ of its members are associated with a ``real group" (or a mock group) defined by the DM halos in the simulation. Conversely, if a``real group" is a complete group, more than $50\%$ of its members are associated with a ``reconstructed group". In addition, if a PFOF group is a pure group but its associated real group is not a complete group, and vice versa, the association is called a one-way match. If a PFOF group is a pure group and its associated real group is also a complete group, the association is called a two-way match. The purity and completeness of a group catalog can be defined according to the association as follows. For one-way match,
\begin{equation}\label{p1}
    p_1 = \frac{N_{\rm{pure}}}{N_{\rm{PFOF}}},
\end{equation}
and
\begin{equation}\label{c1}
    c_1 = \frac{N_{\rm{complete}}}{N_{\rm{real}}},
\end{equation}
where $N_{\rm{pure}}$, $N_{\rm{complete}}$, $N_{\rm{PFOF}}$, and $N_{\rm{real}}$ are the number of pure groups, complete groups, PFOF groups (or reconstructed groups), and real groups, respectively.
For a two-way match,
\begin{equation}\label{p2}
    p_2 = \frac{N_{2}}{N_{\rm{PFOF}}}
\end{equation}
and
\begin{equation}\label{c2}
    c_2 = \frac{N_{2}}{N_{\rm{real}}},
\end{equation}
where $N_{2}$, is the number of pure and complete groups.
Moreover, when a real group is identified as several smaller groups in the reconstructed catalog, the situation is called ``fragmentation". Conversely, overmerging is when two or more real groups are identified as a single reconstructed object.

\begin{figure}
 \begin{center}
  \includegraphics[width=9cm]{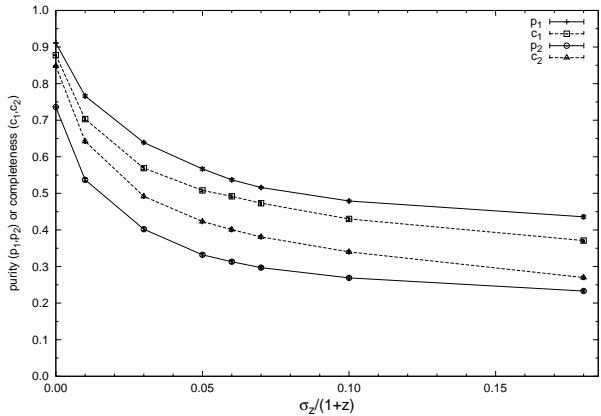}\\
  \caption{Purity or completeness is plotted as a function of photometric redshift accuracy $\sigma_{\Delta z/(1+z_s)}$. In the range of $\sigma_{\Delta z/(1+z_s)}$ between 0.03 and 0.07, purity or completeness drops by less than 20$\%$. For current PS1 Medium Deep data, whose $\sigma_{\Delta z/(1+z_s)}$ is expected to be in the range between 0.03 and 0.07, the performance of the PFOF should not change dramatically. Errorbars show the deviation over 50 realizations of the redshift errors.
   }\label{fig:pcSig}
 \end{center}
\end{figure}

\begin{figure*}
 \begin{center}
  \includegraphics[width=15cm]{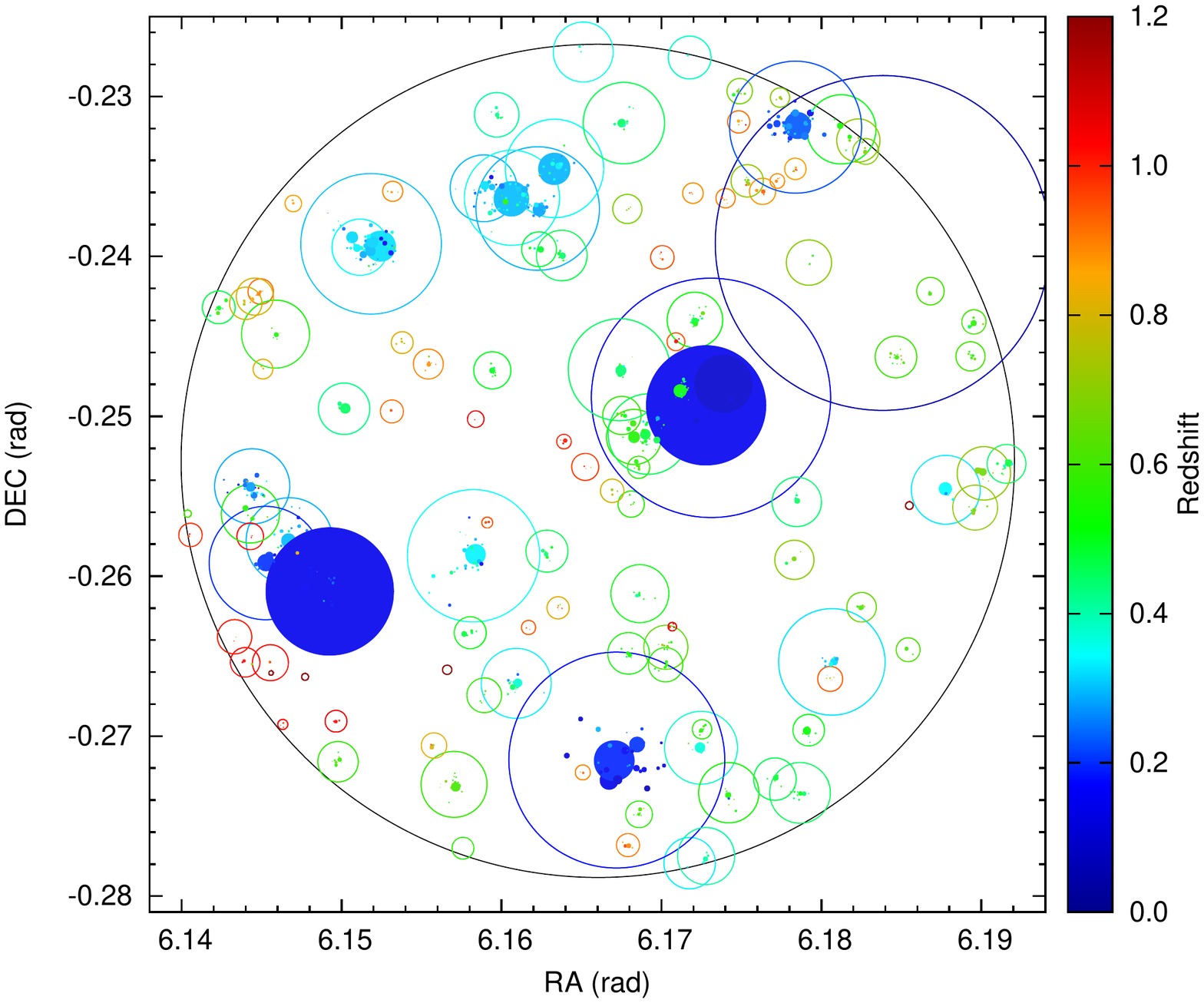}\\
  \caption{Distribution of PFOF and mock groups with halo mass $M_h$ $\geq$ $10^{14} \Msun$ in a simulated case with $\sigma_{\Delta z/(1+z_s)}$ = 0.06. PFOF and mock groups with color-coded redshift are represented with solid and open circles, respectively, and with radii equal to the maximum center-to-member distance. The big black circle denotes the field of view of 7 deg$^2$, equivalent to a single pointing of PS1. It is seen that most of the mock groups are detected by PFOF groups ($\sim$ 97$\%$) with roughly correct redshift and PFOF groups (solid circles) tend to have smaller size and be fragments of the corresponding mock groups.
   }\label{fig:7SqrDegMap}
 \end{center}
\end{figure*}

In practice, a perfect reconstructed group catalog is not achievable, with purity and completeness having a tendency to be mutually exclusive \citep[e.g.][]{ger05,kno09}. A similar tension exists between overmerging and fragmentation. Some statistics introduced by \cite{kno09} and \cite{kno12} to measure ``goodness" in such a way that maximizes (or minimizes) them yields a sort of ``optimal" group catalog. In this study, we adopt three statistics from these works, $g_1$, $g_2$, and $\widetilde{g_1}$, for the purpose of our group catalog optimization. We briefly review these quantities here. The first quantity $g_1$ is defined as
\begin{equation}\label{g1}
    g_1 = \sqrt{\frac{(1-p_1)^2 + (1-c_1)^2}{2}},
\end{equation}
which is normalized to be between 0 and 1. This is slightly different from the original $\texttt{g}_1$ definition in \cite{kno09}, and gives the distance to this optimal point in the $c_1$-$p_1$ plane and thus is a measure of the balance between completeness and purity. The second quantity $g_2$ is defined as
\begin{equation}\label{g2}
    g_2 = \frac{c_2}{c_1} \frac{p_2}{p_1},
\end{equation}
and measures the balance between overmerging and fragmentation. The last quantity $\widetilde{g_1}$ has the form
\begin{equation}\label{g1T}
    \widetilde{g_1} = \sqrt{\frac{(1-p_2)^2 + (1-c_2)^2}{2}},
\end{equation}
which is similar to $g_1$ except that all one-way match statistics are replaced by their two-way match statistic counterparts. For a perfect group catalog, we expect that c1 $\simeq$ c2 $\simeq$ 1 and p1 $\simeq$ p2 $\simeq$ 1, meaning that essentially neither overmerging nor fragmentation is present in the catalog. Therefore, $g_1$ and $\widetilde{g_1}$ should approach 0 and $g_2$ should approach 1.

Since $g_1$ is only based on one-way match statistics, the resulting catalog, if optimized, might contain, unnecessarily, many such overmerged or over-fragmented groups that will exhibit very good one-way statistics but very poor two-way statistics \citep{kno12}. Instead of using $g_1$, we, following \cite{kno12}, take $\widetilde{g_1}$ as the main optimization measure throughout this paper.

\begin{figure}
 \begin{center}
  \includegraphics[width=14cm]{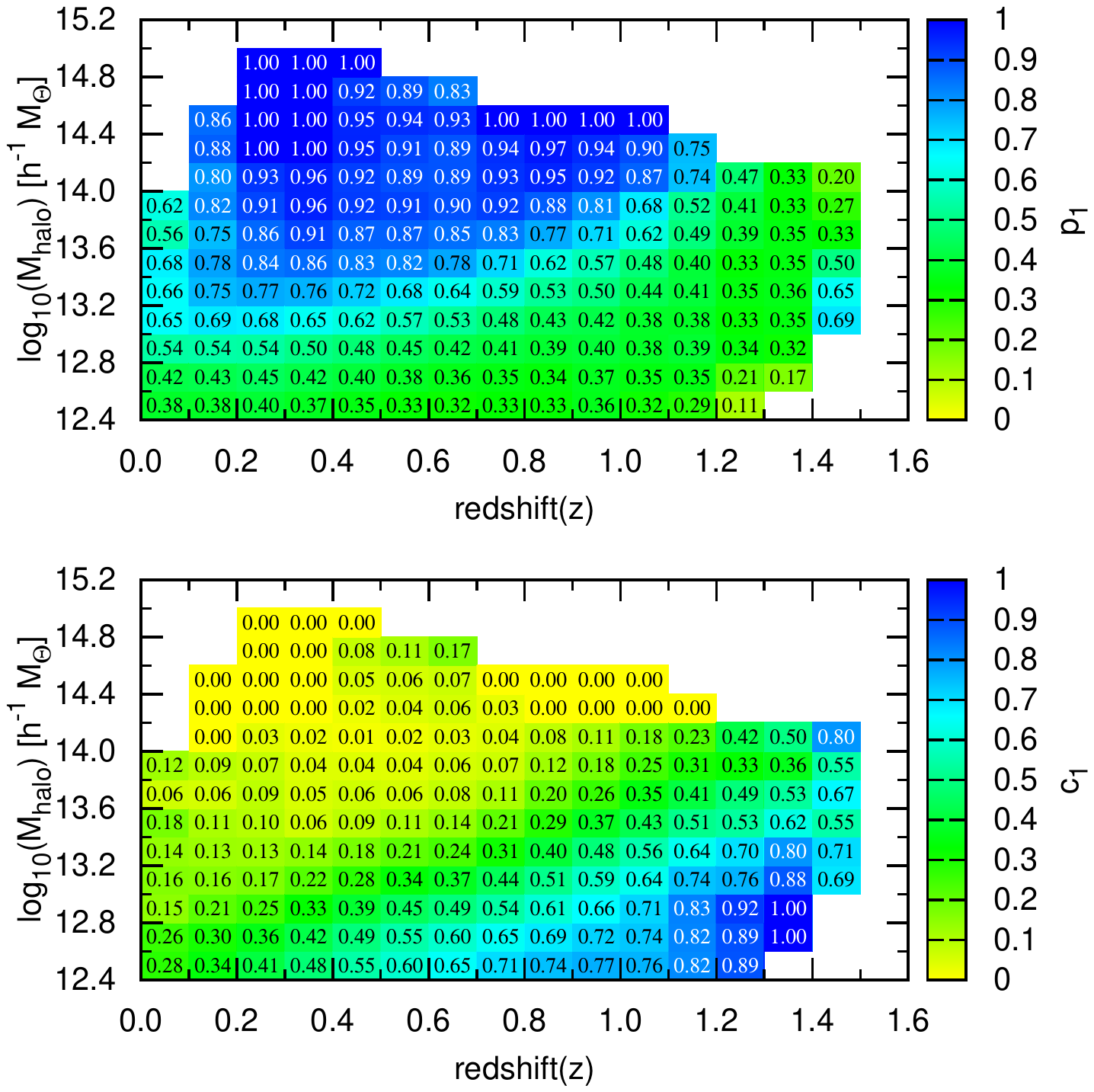}\\
  \caption{ Color-coded purity $p_1$ (top) and completeness $c_1$ (bottom) are plotted as a function of redshift $z$ and halo mass $M_h$. This plot is an illustration of the grouping result with parameters optimized for the the low richness groups. In the high mass range $\sim$ $10^{14} \Msun$, PFOF groups are pure and their corresponding mock groups are incomplete, consistent with what we see in Figure~\ref{fig:7SqrDegMap}. By contrast, in lower mass halos ($log_{10}{M_h}$ $<$ 13.2) and higher redshift ($z$ $>$ 0.8) range, the opposite result is obtained in that PFOF groups are not pure and their corresponding mock groups are more complete. The low completeness for high mass halos is mainly due to optimization over all $N$ $\geq$ 4 groups, which dominate the global measures.
   }\label{fig:p1c1}
 \end{center}
\end{figure}

\subsection{Subset Optimization}
In addition to making use of Durham mock to optimize PFOF grouping, we alternatively optimize our linking lengths and threshold probability, i.e. $l_p$, $l_z$, and $P_{th}$, by utilizing published spectral-$z$ group catalogs as the training set. That is, zCOSMOS groups are the training set for PS1MD04 and DEEP2 groups for PS1MD07. In other words, we use spectral-$z$ groups with a high sampling rate in the same field as the training set so as to avoid any dependence on the semi-analytical model, which may turn out not to be a precise representation of the observational data. The optimization procedure is as follows. For a given set of linking lengths and threshold, PFOF detects photo-$z$ groups from a photometric galaxy sample. For a PFOF group, it may include galaxies with or without a spectroscopic redshift. We identify galaxies with spectral-$z$ in the group to form a sub-group and then evaluate purity and completeness of the sub-group referenced to published spectral-$z$ groups to obtain the optimization measure $\widetilde{g_1}$. We survey a wide range of linking lengths and thresholds to locate the minimum value of $\widetilde{g_1}$ from the sub-sample as our optimization target. The main concern for this method is whether the optimal set of linking lengths and threshold found from these sub-groups is also the optimal one for the full dataset. To examine the feasibility of the method, we then use Durham mock catalog to simulate a sample with a 50$\%$ sampling rate similar to the zCOSMOS or DEEP2 surveys as the sub-sample, and test whether the two minimum $\widetilde{g_1}$'s from the sub-sample and from the full-sample, have the same linking length and threshold.

\begin{figure}
 \begin{center}
  \includegraphics[width=14cm]{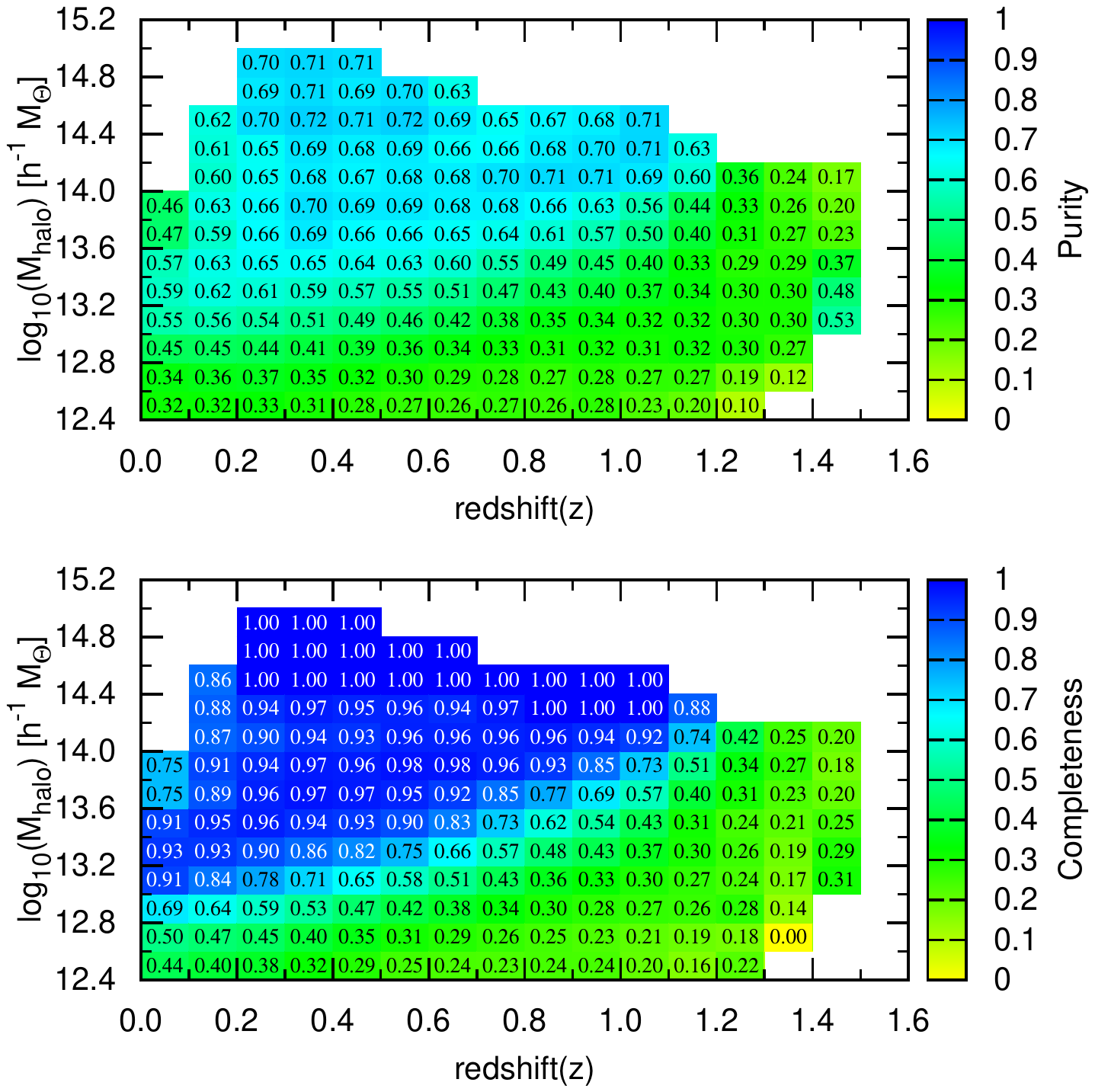}\\
  \caption{Purity (top) and completeness (bottom) computed as defined by \cite{mur12} are displayed as function of $z$ and $M_h$ to illustrate that for the same PFOF group catalog, depending on the definition of purity or completeness used, the PFOF performance can change appreciably.
   }\label{fig:pcORCA}
 \end{center}
\end{figure}

\begin{figure}
 \begin{center}
  \includegraphics[width=14cm]{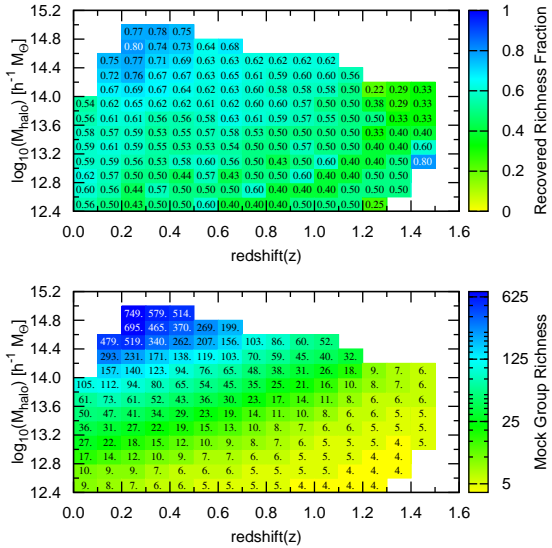}\\
  \caption{Recovered richness fraction from PFOF groups (top) and original richness from their corresponding mock groups (bottom) are shown in terms of $z$ and $M_h$. PFOF groups compose $\gtrsim$ 60$\%$ of members from their original mock groups with high richness, $N$ $\geq$ 10. For a fixed richness, the distribution of the richness is a linear function of $z$ and $M_h$, and this is due to a limiting magnitude selection for the sample.
  }\label{fig:richness}
 \end{center}
\end{figure}

\begin{figure*}
 \begin{center}
 \epsscale{1.0}
  \plotone{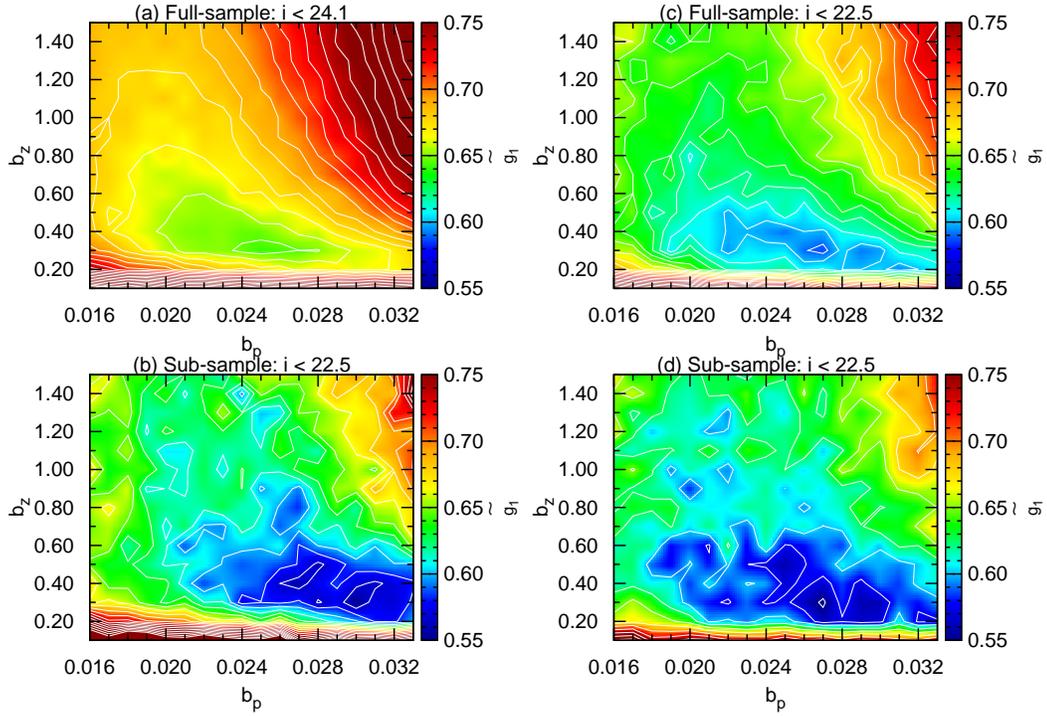}
  \caption{The optimization measure $\widetilde{g_1}$ is plotted as a function of linking length parameter parallel ($b_z$) and perpendicular ($b_p$) to the line-of-sight at a fixed threshold, $p_{th}$. To simulate the survey sampling rate and area of zCOSMOS, we randomly select 50$\%$ of galaxies from the Durham mock in 2 square degree as the sub-sample. In addition, we consider one case with the same magnitude cut (i $\leq$ 22.5) for both full (c) and sub-sample (d), and the other one with a different magnitude cut applied to the full (a) (i $\leq$ 24.1) and sub-sample (b) (i $\leq$ 22.5). In both two cases, the simulated photo-$z$ accuracy 0.06 is applied. It is found that in both cases the minima of $\widetilde{g_1}$ from full and sub-sample coincide with each other, i.e. the optimal set of linking lengths and threshold selected from the sub-sample is also the same one from the full-sample. Therefore, subset optimization is feasible.
   }\label{fig:SubsetCos}
 \end{center}
\end{figure*}

\begin{figure}
 \begin{center}
 \epsscale{1.7}
  \plotone{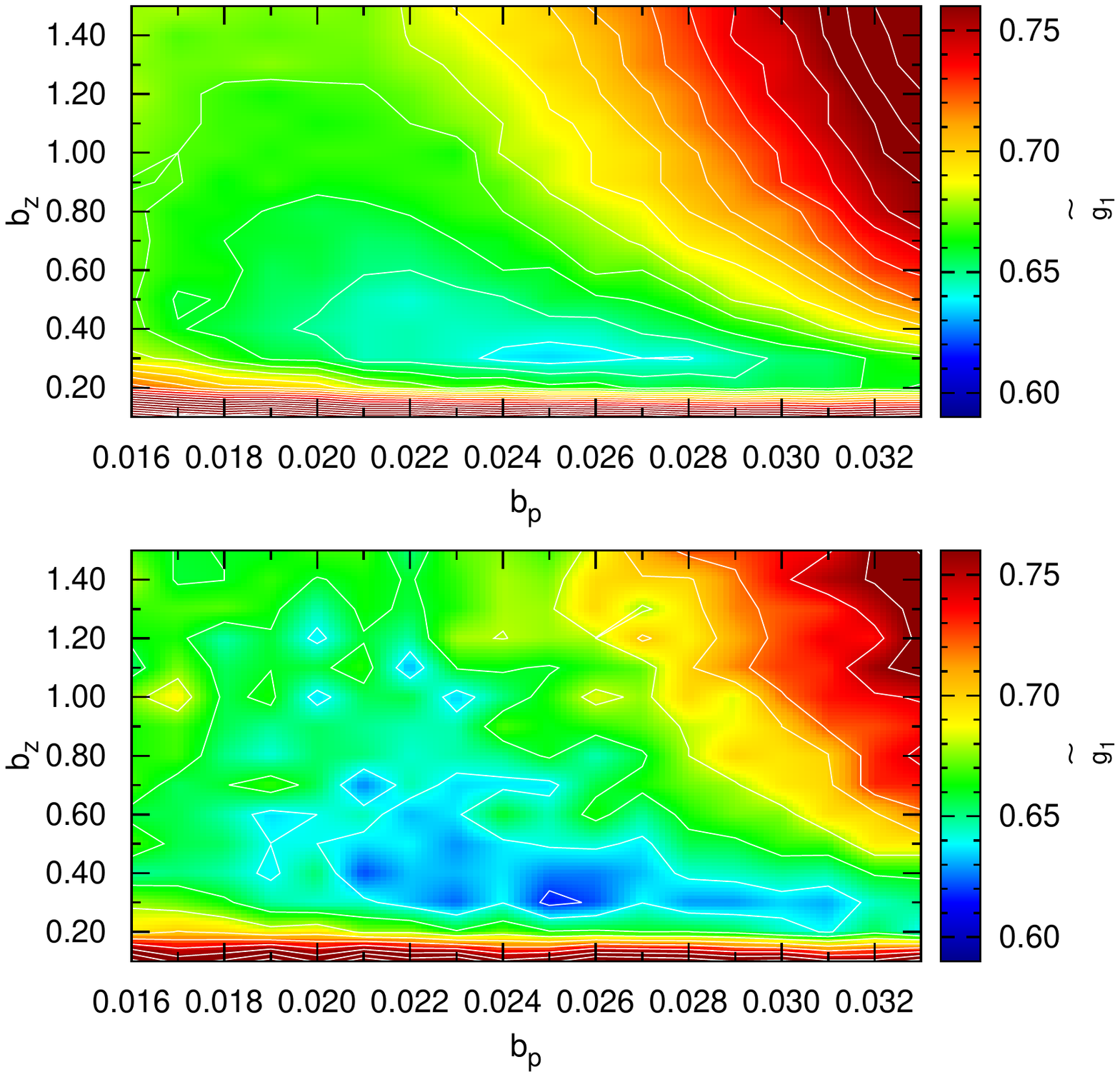}
  \caption{Similar to Figure~\ref{fig:SubsetCos}. We simulate the survey sampling rate ($\sim$ 50$\%$) and area ($\sim$ 1 deg$^{2}$) similar to the DEEP2 EGS using the Durham mock. We again find that the optimal set of linking lengths and threshold with the minimum $\widetilde{g_1}$ found from the sub-sample are the same as those derived from the full-sample.
   }\label{fig:SubsetEGS}
 \end{center}
\end{figure}

\section{PS1 MOCK GROUP CATALOG}
For a FOF based grouping algorithm, linking length calibration is important and necessary. Mocks, in spite of their possibly inaccurate realization of the Universe, can provide not only self-consistent tests but can also be the tool to evaluate our training efficiency.

\subsection{Performance Dependence on Photometric Redshift Accuracy}
To assess the dependence of the performance of PFOF on photometric redshift (or photo-$z$) accuracy, we apply PFOF to various simulated widths of photo-$z$ error ranging from $\sigma_{\Delta z/(1+z_s)}$ = $10^{-4}$ to 0.2. We select galaxy samples with $i$ $\leq$ 24.1 and $z$ $\leq$ 1.4, and evaluate purity and completeness for groups with richness $N \geq 4$. For the photo-$z$ simulation, we perturb the original observed redshift $z_{obs}$ by adding an extra Gaussian distributed $\delta z$ with variance of a given photo-$z$ error $\sigma_{\Delta z/(1+z_s)}$. Additionally, an outlier rate of $4\%$ is considered in the photo-$z$ simulation in redshift range 0 $<$ $z$ $<$ 1.4. For a detailed discussion of the simulation of photo-$z$, see Liu08.

Fig.~\ref{fig:pcSig} shows the result of the optimal purity and completeness as functions of photo-$z$ error, $\sigma_{\Delta z/(1+z_s)}$. The errorbars show the 1 $\sigma$ deviation for 50 realizations. It can be seen that the purity and completeness drop when the photo-$z$ error increases. This is expected since a large uncertainty in redshift makes group recovery more unreliable. In addition, the small errorbars also indicate the stability of PFOF algorithm. Varying $\sigma_{\Delta z/(1+z_s)}$ = 0.03 to 0.07, $p_1$ and $c_1$ decline by $\sim$ 19$\%$ and $\sim$ 14$\%$, respectively. Over this range of errors, the performance of PFOF does not change significantly. That is, PFOF performance is not sensitive to photo-$z$ accuracy for a PS1 MDS like survey where the expected photo-$z$ accuracy falls into this range.

\begin{figure*}
 \begin{center}
 \epsscale{1.0}
  \plotone{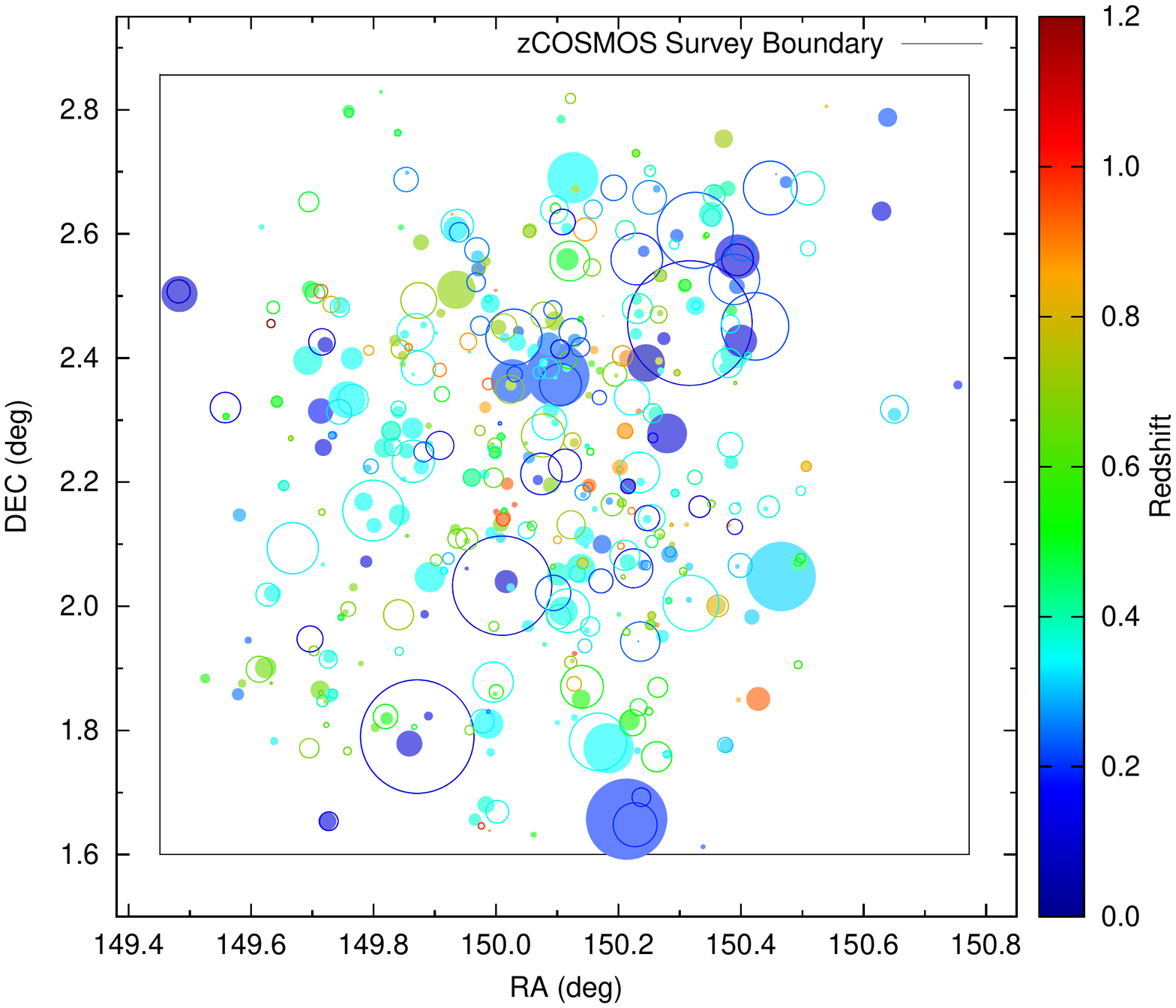}
  \caption{ Distribution of PFOF COSMOS subgroups (solid) and zCOSMOS groups (open) with color-coded redshift and radius equal to maximum center-to-member distance in the redshift range 0 $<$ z $<$ 1.2. We obtain the optimal values of $p_1$ = 66$\%$ and $c_1$ = 70$\%$.
   }\label{fig:COSMOS}
 \end{center}
\end{figure*}

\begin{figure}
 \begin{center}
 \epsscale{1.4}
  \plotone{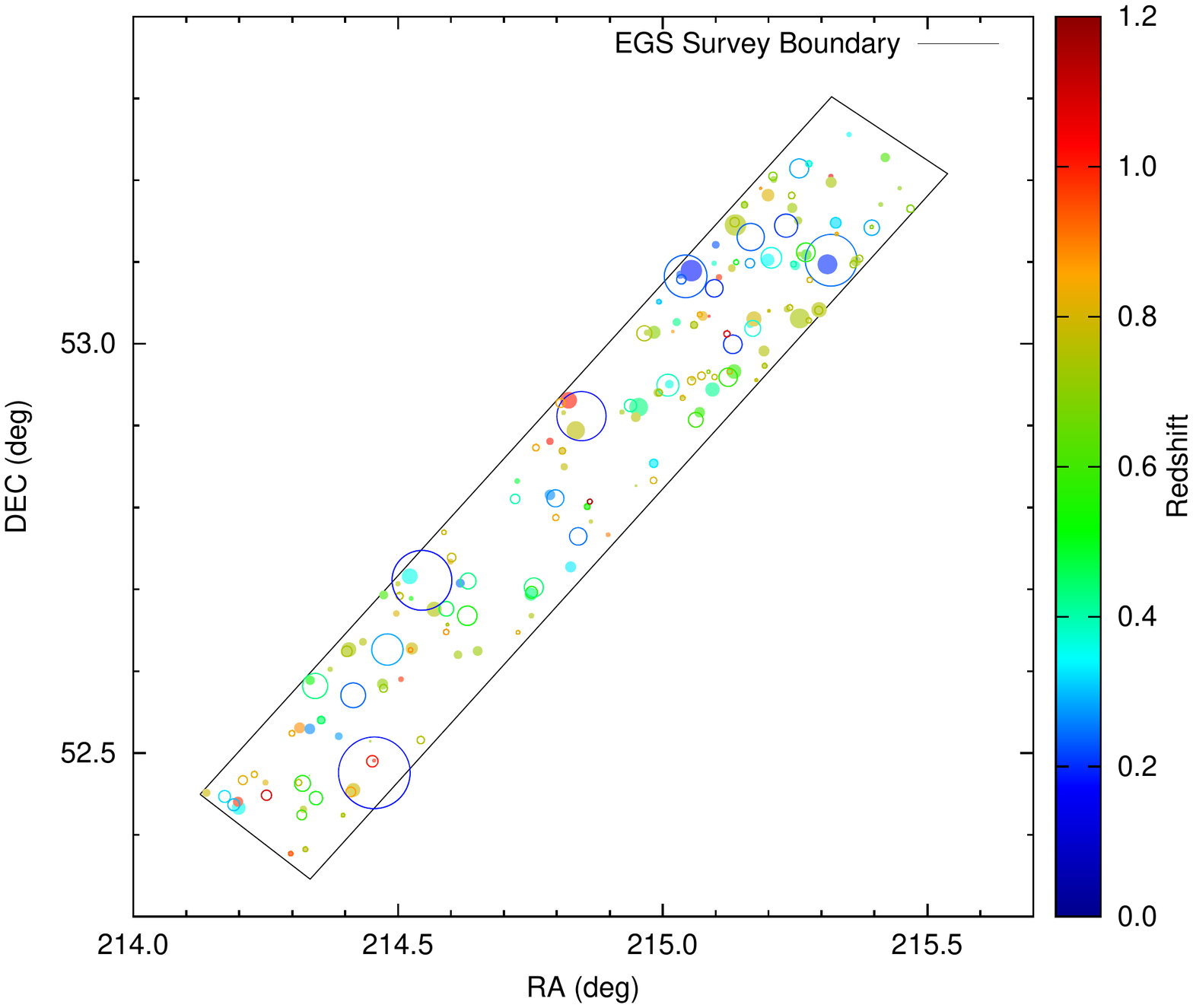}
  \caption{Distribution of PFOF EGS subgroups (solid) and DEEP2 EGS groups (open) with color-coded redshift and radius equal to maximum center-to-member distance. For this case, we obtain the optimal values of $p_1$ = 55$\%$ and $c_1$ = 63$\%$ from the sub-sample.
   }\label{fig:EGS}
 \end{center}
\end{figure}

\subsection{The Mock Group Catalog with $\sigma_{\Delta z/(1+z_s)}$ = 0.06}
The typical error width of photo-$z$ accuracy $\sigma_{\Delta z/(1+z_s)}$, for five-band surveys, such as Pan-STARRS Medium Deep Survey fields, is roughly 0.06 \citep{sag12}. We shall illustrate the PFOF group finding results with this error using the mock catalog as follows. The selection cut for galaxies is the same as in the previous test and we focus on groups with richness $N$ $\geq$ 4. The optimal values of $p_1$ and $c_1$ are found to be 54$\%$ and 49$\%$, respectively, after optimization.

Figure~\ref{fig:7SqrDegMap} shows a simple comparison between the PFOF and mock group catalogs by plotting both sets of clusters residing in haloes mass $M_h$ $\geq$ $10^{14}$ $\Msun$ up to $z$=1.4. Centers of open and solid circles with color-coded redshift represent the positions of the mock and PFOF groups, and their radii give the maximum group member distance from the group center. The big black circle gives the field of view of 7 deg$^2$, equivalent to a single pointing of PS1. It can be seen that most mock groups are detected by PFOF (112 out of 116), where the successful detection is to recover a minimum of 5 member galaxies from the mock groups, and these PFOF groups are at roughly the correct redshift. But the mock groups appear to be fragmented by the PFOF detection. To quantify the grouping performance, we follow the work of \cite{mur12} and  plot the color-coded purity and completeness are plotted as functions of dark matter halo mass, $M_h$, and redshift $z$, with sampling intervals of 0.05 in redshift and 0.2 in log$_{10}$ halo mass. We also smooth the data using a 3 $\times$ 3 grid in redshift and log mass bins, for which the purity or completeness of a given cell is the mean value over this coarse region, and set a threshold of at least five clusters detected in this region. In Figure~\ref{fig:p1c1}, $p_1$ (top) and $c_1$ (bottom) are shown respectively. It can be seen that for high masses ($M_h$ $\geq$ $10^{14} \Msun$), the recovered groups are highly pure but not compete, consistent with the result from Figure~\ref{fig:7SqrDegMap}. The opposite trend can be seen between purity and completeness, i.e. the highly complete region has low purity. Note that the high purity and low completeness in the high halo mass region seem to disagree with the optimal values of $p_1$ and $c_1$, being equal to 54$\%$ and 49$\%$, respectively. This is mainly due to the fact that the optimal $p_1$ and $c_1$ are global measures, derived by optimization over all $N$ $\geq$ 4 groups; the abundant low $N$ groups dominate the global measures and compromise the high-mass-end results. 

\begin{figure*}
 \begin{center}
 \epsscale{1.0}
  \plotone{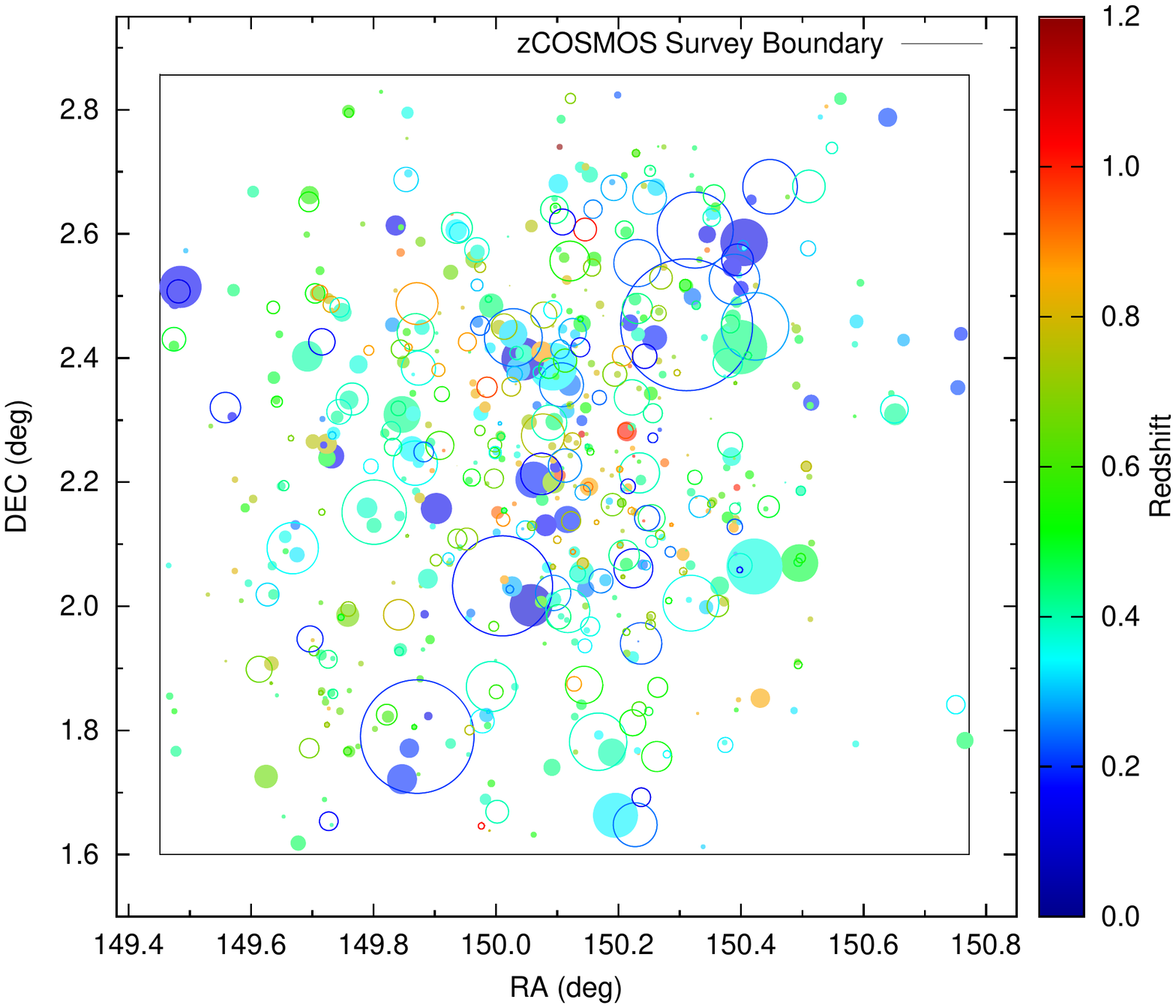}
  \caption{Similar to Figure~\ref{fig:COSMOS}, we plot the distribution of PFOF PS1MD04 subgroups (solid) and zCOSMOS groups (open) with color-coded redshift and radius equal to maximum center-to-member distance. For this case, the optimal values of $p_1$ = 40$\%$ and 57$\%$ are obtained.
   }\label{fig:PS1MD04}
 \end{center}
\end{figure*}

\begin{figure}
 \begin{center}
  \includegraphics[width=10cm]{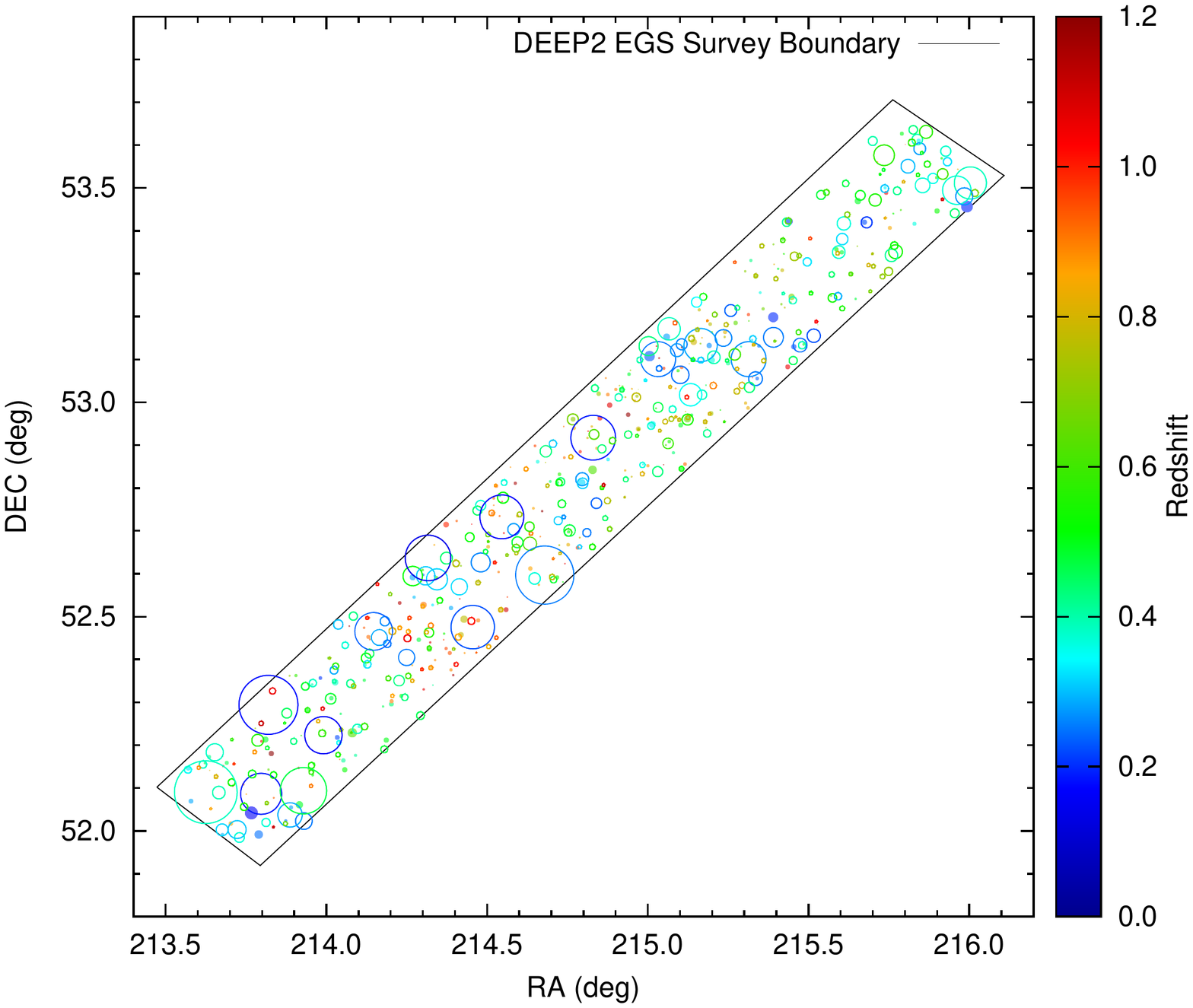}\\
  \caption{Similar to Figure~\ref{fig:EGS}, distribution of PFOF PS1MD07 subgroups (solid) and DEEP2 EGS groups (open) is plotted with color-coded redshift and radius equal to maximum center-to-member distance. In this case, we get the optimal values of $p_1$ = 53$\%$ and $c_1$ = 47$\%$.
   }\label{fig:PS1MD07}
 \end{center}
\end{figure}

In addition, we note that performance based on purity and completeness can vary significantly with their definitions. In the following, we adopt the definitions of purity and completeness from the ORCA method of \cite{mur12} to illustrate the dependence of performance on the definitions. For an ORCA complete group, a halo is detected if at least $N_{min}$ galaxies are identified, even if they are shared between multiple ORCA clusters. In ORCA, purity is defined as the fraction of galaxies assigned to the group that are members of the host halo. With the same group catalog used previously for $p_1$ and $c_1$ in Figure~\ref{fig:p1c1}, we plot purity (top) and completeness (bottom) with the ORCA definition in Figure~\ref{fig:pcORCA} to demonstrate the difference. For the purity comparison, the difference is small. This is simply because two purity definitions are close to each other. ORCA purity drops slightly compared to PFOF purity. However, for completeness, the discrepancy between the two definitions is large. The original low completeness in high mass region in PFOF definition becomes high completeness with ORCA definition while the original high completeness in the low mass region using the PFOF definition becomes lower completeness in ORCA definition.

Finally, we address the richness recovery. Richness of mock (bottom) and recovered richness fraction of PFOF (top) groups is plotted in terms of halo mass $M_h$ and redshift $z$ for comparison in Figure~\ref{fig:richness}. The recovered richness is defined as the richness sum of PFOF groups corresponding to the same mock group, and the correspondence is assured by the one-way match criterion. It is found that PFOF can detect $\gtrsim$ 60$\%$ of the members from high richness groups, $N$ $\geq$ 10. In other words, at a specified $M_h$ and $z$, the recovered richness is $\sim$ 25$\%$ less than mock group richness. Additionally, for the same richness, groups are more massive at higher redshift due to the applied limiting magnitude selection.

\subsection{Depth and Absolute Magnitude Selection}
To understand how the depth of a sample affects our group finder performance, we select a sample with $i$ $\leq$ 25.8 and a simulated photo-$z$ width $\sigma_{\Delta z/(1+z_s)}$ $\sim$ 0.06 in the redshift range between 0 and 1.4 for PFOF grouping. We obtain optimal values of $p_1$ = 51$\%$ and $c_1$ = 45$\%$, which are $\sim$ 8$\%$ lower than those found for a sample with a shallower depth $i$ $\leq$ 24.1. It appears that for PFOF grouping, increasing depth can lead to a worse performance. However, the effect is not significant.

Similarly, a sample based on a rest frame magnitude selection is also setup to probe the group finding performance. A sample is selected with $z$ up to 1.4 with an absolute magnitude cut of $i$ $\leq$ -19.0, corresponding to roughly the similar galaxy number density as to that in the sample with flux-limit $i$ $\leq$ 24.1, plus a simulated photo-$z$ width $\sigma_{\Delta z/(1+z_s)}$ $\sim$ 0.06 to search for the optimal grouping performance. The optimal values of $p_1$ and $c_1$ are 62$\%$ and 61$\%$, respectively. Compared to that of the flux-limit sample, the performance has a roughly 20$\%$ increase.

\subsection{Subset Optimization Study} \label{subset}
In this section, we make use of the Durham mock catalog to setup two cases for studies of subset optimization. In the first case, we select $50\%$ of galaxies with $i$ $\leq$ 22.5 in a 2-deg$^{2}$ field to be our subsample to mimic the roughly $50\%$ sampling completeness case, e.g. 20k zCOSMOS galaxies in PS1MD04. For the full sample, two different photometric depths, $i$ $\leq$ 24.1 and $i$ $\leq$ 22.5, are applied. When we look for optimal linking lengths and threshold, we find that the minimal $\tilde{g_1}$, the optimization measure, is located at $P_{th}$ = 0.001. To illustrate how the optimization measure evolves with linking length, we plot the color-coded $\tilde{g_1}$ at fixed $P_{th}$ = 0.001 as a function of linking parameters perpendicular and parallel to the line of sight, $b_p$ and $b_z$, over-plotted with $\tilde{g_1}$ contours in Figure~\ref{fig:SubsetCos}. The left two panels shows results for the full sample (upper) with depth $i$ $\leq$ 24.1 (a) and for the sub-sample (lower) with $i$ $\leq$ 22.5 (b), while the right two panels are also for the full sample (upper) (c) and for the sub-sample (lower) (d) but both with depth $i$ $\leq$ 22.5. That is, there is a depth difference between the full and sub-sample (the left two panels) but no depth difference between the right two panels. It can be seen that from the left two panels, there are two local minimal $\tilde{g_1}$'s in the sub-sample. One matches the minimum in the full sample, but the other deviates from that in the full sample which turns out to be the global minimum. Therefore, if we identify the optimal linking lengths using a sub-sample with a certain depth difference from its full sample, it seems that the shorter optimal linking lengths are likely to be the correct optimal linking lengths. On the other hand, from the right panels the two minima are located at roughly the same linking lengths, meaning that we can find the real optimal linking lengths for the full sample via ``subset optimization".

In the other case, we select $50\%$ of galaxies with $i$ $\leq$ 24.1 in a 1-deg$^{2}$ field to be our sub-sample to mimic roughly $50\%$ of sampling completeness for DEEP2 EGS galaxies in PS1MD07, and the same depth as in the sub-sample is set for the full sample. Similar to Figure~\ref{fig:SubsetCos}, we also plot the optimization measure $\tilde{g_1}$ at fixed $P_{th}$ = 0.001 in Figure~\ref{fig:SubsetEGS}. The result shows that the minimum in the sub-sample can match that in the full sample at the same linking lengths. We thus conclude that the survey area of the sub-sample appears to be not a relevant factor.

From above tests, we find that a shallower depth in a sub-sample than in a full-sample does not affect the optimization of PFOF. In other words, for the purpose of training, a spectroscopic survey with a shallow depth is enough to support the PFOF optimization. On the other hand, from the observational point of view, a shallow spectroscopic survey is easier to achieve than a deep one. This makes the optimization of PFOF easy to apply to observation data.

\section{Observational Data Applications}
In this section, we utilize four observational datasets, including photometric galaxy catalogs of COSMOS, EGS, PS1MD04, and PS1MD07, to illustrate the grouping performance of PFOF via subset optimization for various photo-$z$ error widths, $\sigma_{\Delta z/(1+z_s)}$ = $\sim$ 0.01 (COSMOS), 0.03 (EGS), and 0.06 (PS1MD04 and PS1MD07), respectively. For PFOF grouping, the minimum richness cut for a sub-sample and full-sample are 3 and 4, respectively.

\subsection{COSMOS Group Catalog}
We make use of public COSMOS galaxy catalog \citep{geo11} for PFOF group finding. In this catalog, the typical redshift accuracy can reach $\sigma_{\Delta z/(1+z_s)}$ $\lesssim$ 0.007 for galaxies with F814W $<$ 22.5, and $\sigma_{\Delta z/(1+z_s)}$ = 0.012 for F814W = 24, at z $<$ 1.2 \citep{ilb09}. The catalog contains 115,831 galaxies with F814W $<$ 24.2 up to $z$ $\sim$ 4.6, and the survey area is $\sim$ 2-deg$^2$. We focus on galaxies below redshift 1.4. In total we obtain 104,060 galaxies in the full sample. In addition, we take the zCOSMOS 10k group galaxies \citep{kno09} to be our sub-sample. There are 11,262 zCOSMOS galaxies with $z$ $<$ 1.4 and i $\leq$ 22.5 in the area of 2-deg$^2$.

Via the subset optimization, we obtain overall optimal values of $p_1$ = 66$\%$ and $c_1$ = 70$\%$ from the zCOSMOS sub-sample with the linking probability threshold $P_{th}$ = 0.04 and the redshift dependent comoving linking lengths parallel and perpendicular to line-of-sight, $l_p$ and $l_z$, between 0.10 and 0.55 Mpc for $l_p(z)$ and between 3.4 and 19.06 Mpc for $l_z(z)$. PFOF reconstructs 249 groups in the zCOSMOS sub-sample, which originally had 227 spectral-$z$ groups. On the other hand, PFOF also detects a total of 3,314 groups in the full COSMOS sample, with 20,954 group galaxies ($\sim$ 20.2$\%$ of the original galaxy sample). In Figure~\ref{fig:COSMOS}, we plot PFOF groups in the sub-sample using solid circles and zCOSMOS groups with open circles with color-coded $z$ and radius, the maximum member-group center distance, in the redshift range 0 $<$ $z$ $<$ 1.2. We compute $p_1$ and $c_1$ in six redshift bins, (a) 0 $<$ $z$ $<$ 0.2, (b) 0.2 $<$ $z$ $<$ 0.4, (c) 0.4 $<$ $z$ $<$ 0.6, (d) 0.6 $<$ $z$ $<$ 0.8, (e) 0.8 $<$ $z$ $<$ 1.0, and (f) 1.0 $<$ $z$ $<$ 1.2. We find that in (a) $p_1$ = 14/21 $\approx$ 67$\%$ and $c_1$ = 12/22 $\approx$ 55$\%$, in (b) $p_1$ = 76/119 $\approx$ 64$\%$ and $c_1$ = 62/93 $\approx$ 67$\%$, in (c) $p_1$ = 28/36 $\approx$ 78$\%$ and $c_1$ = 31/38 $\approx$ 82$\%$, in (d) $p_1$ = 32/49 $\approx$ 65$\%$ and $c_1$ = 38/52 $\approx$ 73$\%$, in (e) $p_1$ = 14/24 $\approx$ 58$\%$ and $c_1$ = 14/21 $\approx$ 67$\%$, and in (f) no group was found in this bin. We also can observe some fragmented and over-merged subgroups from PFOF detections but the fragmentation and overmerger problem are not serious.

To assess the performance of membership identification from PFOF, we compare our group galaxies of COSMOS to X-ray membership galaxies from \cite{geo11}. We find that the photo-$z$ dispersion from PFOF group galaxies are slightly smaller than that from X-ray cluster galaxies. PFOF performance thus is comparable to that of other group finders.

\subsection{EGS Group Catalog}\label{EGS}
In the EGS photometric galaxy sample, we have the same selection cut for galaxies, with $i$ $\leq$ 24.1 and photo-$z$ up to 1.4, and obtain 8,558 galaxies and 3,526 galaxies to be the input full-sample and sub-sample, respectively. After optimization, we find that the optimal values of $p_1$ and $c_1$ are 55$\%$ and 63$\%$ from the sub-sample with $p_{th}$ = 0.01, $l_p$$(z)$ = 0.282 to 0.560 Mpc, and  $l_z$$(z)$ = 3.41 to 6.87 Mpc. In the sub-sample, there are 96 spectral-$z$ groups, and PFOF identifies 69 sub-groups while in EGS full-sample, 223 groups ($N$ $\geq$ 4) are detected and they contain 1,237 group galaxies ($\sim$ 14$\%$ of EGS sample). Similar to Figure~\ref{fig:COSMOS} but without redshift binning, the location, color-coded redshift, and radius of 98 DEEP-EGS (solid circle) and 107 PFOF (open circle) sub-groups are plotted in Figure~\ref{fig:EGS}. In 6 redshift ranges of equal partition between $z$ = 0 and 1.2, we find that (a) $p_1$ = 0/1 = 0. and $c_1$ = 1/5 = 20$\%$, (b) $p_1$ = 13/21 $\approx$ 62$\%$ and $c_1$ = 13/24 $\approx$ 54$\%$, (c) $p_1$ = 8/14 = 57$\%$ and $c_1$ = 10/17 $\approx$ 59$\%$, (d) $p_1$ = 29/51 $\approx$ 57$\%$ and $c_1$ = 27/35 $\approx$ 77$\%$, (e) $p_1$ = 8/16 = 50$\%$ and $c_1$ = 9/12 = 75$\%$, and (f) $p_1$ = 1/4 = 25$\%$ and $c_1$ = 2/5 = 40$\%$.

\subsection{PS1MD04 Group Catalog}
In PS1MD04, we again utilize zCOSMOS 10k groups as the sub-sample for subset optimization. We apply the same selection cut, where $\ips$ $\leq$ 24.1 and photo-$z$ $\leq$ 1.4, to the PS1MD04 catalog, and acquire 345,446 and 9,557 galaxies for the full and sub-sample, respectively. The optimal values of $p_1$ and $c_1$ from the sub-sample are found to be 40$\%$ and 57$\%$ with $p_{th}$ = 0.001, $l_p$$(z)$ = 0.16 to 0.52 Mpc, and  $l_z$$(z)$ = 0.51 to 1.64 Mpc. We identify 427 PFOF subgroups from the zCOSMOS sub-sample which has 227 groups, and in total detect 15,787 groups in PS1MD04 and 108,779 group galaxies, roughly 31.5$\%$ of PS1MD04 sample. In Figure~\ref{fig:PS1MD04}, we plot the result of our subgroup finding by splitting it into 6 redshift bins as we did in Figure~\ref{fig:COSMOS}. In (a) $p_1$ = 12/26 $\approx$ 46$\%$ and $c_1$ = 7/15 $\approx$ 47$\%$, in (b) $p_1$ = 40/111 $\approx$ 36$\%$ and $c_1$ = 39/78 $\approx$ 50$\%$, in (c) $p_1$ = 53/136 $\approx$ 39$\%$ and $c_1$ = 40/60 $\approx$ 66$\%$, in (d) $p_1$ = 54/105 $\approx$ 51$\%$ and $c_1$ = 35/53 $\approx$ 66$\%$, in (e) $p_1$ = 8/44 $\approx$ 18$\%$ and $c_1$ = 7/19 $\approx$ 37$\%$, and in (f) $p_1$ = 2/4 = 50$\%$ and $c_1$ = 1/2 = 50$\%$.

\begin{figure}
 \begin{center}
  \includegraphics[width=15cm]{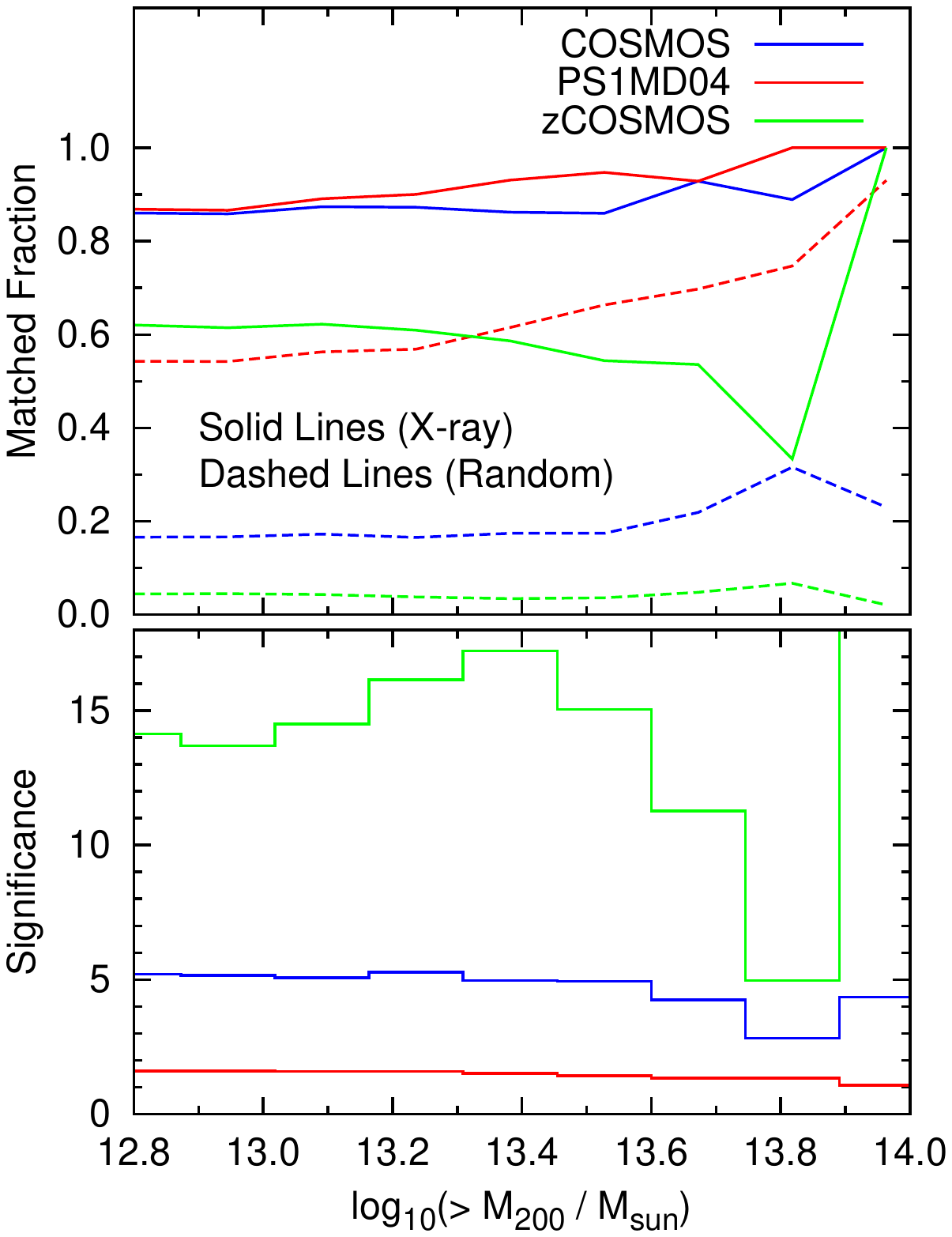}\\
  \caption{The matched fraction (top) and significance (bottom) are plotted as a function of threshold $log_{10}(M_{200}/M_{\bigodot})$ for the case using zCOSMOS groups as the training set, where the significance is defined as the ratio of the matched fraction from the X-ray catalog to the average fraction from the random catalogs. The dashed lines give the matched fraction from random catalogs. It is seen that the matched fraction $\sim$ 85$\%$ for COSMOS and 87$\%$ for PS1MD04. The higher matched fractions from PSMD04 are due to a looser constraint on the redshift difference compared to that from COSMOS, leading to a lower value of significance.
   }\label{fig:match}
 \end{center}
\end{figure}

\subsection{PS1MD07 Group Catalog}\label{PS1MD07}
In PS1MD07, DEEP2-EGS groups are adopted again as the sub-sample. A selection cut is applied, with $\ips$ $\leq$ 24.1 and photo-$z$ $\leq$ 1.4, to the catalog, and we obtain 423,244 and 10,833 galaxies for the two samples. The optimal values of $p_1$ and $c_1$ from the sub-sample are found to be 53$\%$ and 47$\%$ with $p_{th}$ = 0.001, $l_p$$(z)$ = 0.16 to 0.37 Mpc, and  $l_z$$(z)$ = 0.71 to 1.65 Mpc. We reconstruct 305 PFOF subgroups from the DEEP2-EGS sub-sample (originally 309 groups) as plotted in Figure~\ref{fig:PS1MD07}, and in total detect 11,491 groups and 63,044 group galaxies in PS1MD04, roughly 15$\%$ of the PS1MD07 sample. In the 6 redshift ranges, it is found that (a) $p_1$ = 1/3 $\approx$ 33$\%$ and $c_1$ = 0/6 = 0, (b) $p_1$ = 31/40 $\approx$ 78$\%$ and $c_1$ = 26/87 $\approx$ 30$\%$, (c) $p_1$ =31/73 $\approx$ 42$\%$ and $c_1$ = 41/98 $\approx$ 42$\%$, (d) $p_1$ = 38/70 $\approx$ 54$\%$ and $c_1$ = 41/62 $\approx$ 66$\%$, (e) $p_1$ = 43/84 $\approx$ 51$\%$ and $c_1$ = 27/43 $\approx$ 63$\%$, and (f) $p_1$ = 15/34 = 44$\%$ and $c_1$ = 11/13 $\approx$ 85$\%$.

\begin{figure*}
 \begin{center}
  \includegraphics[width=15cm]{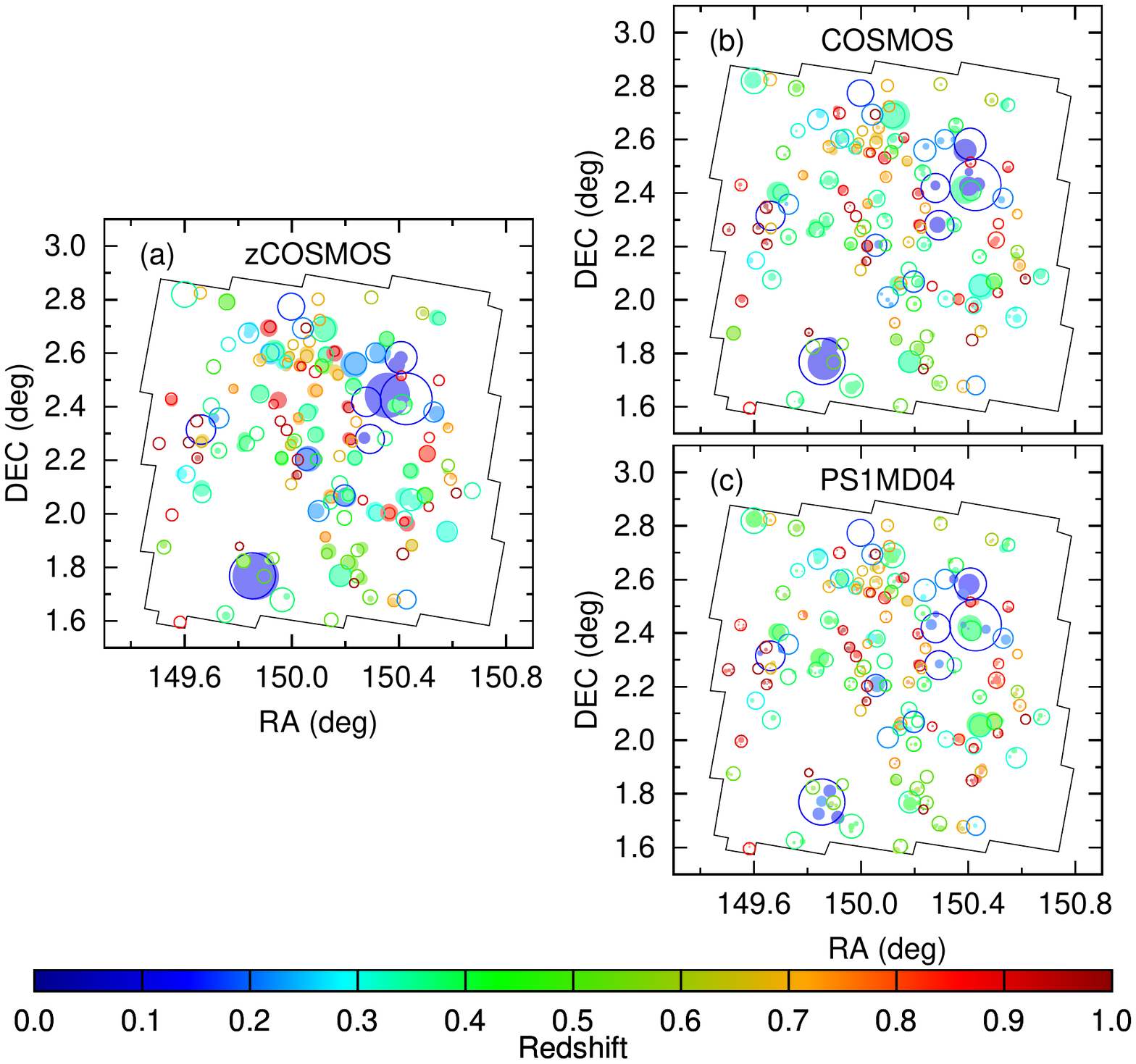}\\
  \caption{Distribution of matched PFOF groups and X-ray clusters is plotted with color-coded redshift and radius equal to maximum center-to-member distance. The black solid line gives the X-ray survey boundary. The open circles denote the X-ray clusters while the solid circles represent the PFOF groups. The radius of circles is equal to the maximum center-to-member distance. The left panel is for zCOSMOS (a), results from COSMOS and PS1MD04 are plotted in (b) and (c), respectively, using zCOSMOS as the training set. Comparing (c) to (b), X-ray clusters tend to be fragmented by PS1MD04 groups more than by COSMOS groups.
   }\label{fig:XrayMatchMap}
 \end{center}
\end{figure*}

\subsection{Comparisons to the X-ray Catalog in the COSMOS field}
To further assess the PFOF performance, we compare our group catalogs in the COSMOS field, i.e. the COSMOS and PS1MD04 catalogs, to the XMM-Newton plus Chandra X-ray catalog \citep{geo11} in the same field, and compute the number of matched PFOF detected groups as a fraction of the number of the X-ray clusters. The criteria for a successful match (or detection) are that the center of a PFOF group has to be inside $r_{200}$, the radius within which the mean density is 200 times the critical density of the universe at the redshift of the group, of an X-ray cluster and that the redshift difference between a PFOF group and an X-ray cluster is within 1.5 $\sigma_{\Delta z/(1+z_s)}$, where $\sigma_{\Delta z/(1+z_s)}$ is the photo-$z$ uncertainty width of a sample. The factor of 1.5 is an adjustable value roughly to account for a probably worse photo-$z$ accuracy in group redshift. In addition, to know whether the match occurs by chance, we construct a random catalog by randomly distributing the position of X-ray clusters but keeping their original redshifts and $r_{200}$, and examine the match between PFOF and the random catalog. We repeat the procedure 50 times and average over the matched fractions, and then define the significance as the ratio of the matched fraction from the X-ray catalog to the average fraction from the random catalogs to understand the effect of coincidence. When the significance is equal to 1, it implies that the match with the X-ray catalog may be totaly caused by chance. On the other hand, when the significance is high, it indicates that the purity of the group catalog is likely also high. In Figure~\ref{fig:match}, we plot the matched fraction (upper) for PS1MD04 (red), COSMOS (blue), and zCOSMOS (green) and significance (lower) as functions of the threshold mass of $M_{200}$ of X-ray clusters using zCOSMOS catalog as the references. From Figure~\ref{fig:match}, it can be seen that the matched fraction of the three catalogs, remains roughly constant as the value of $M_{200}$ decreases, indicating that our detections for low mass X-ray clusters do not decrease significantly. Despite the higher matched fraction $\sim$ 87$\%$ from the PS1MD04 catalog compared to the COSMOS matched fraction ($\sim$ 85$\%$), its matched fraction to the random catalog is also significantly high ($\sim$ 54$\%$), and hence its significance turns out to be the lowest ($\sim$ 1.5), implying that the high photo-$z$ uncertainty (or a looser constraint in redshift difference) causes more X-ray detections by chance. While the matched fraction from COSMOS catalog is $\sim$ 85$\%$ and significance is $\sim$ 5.0 times higher than expected for random matches, the zCOSMOS catalog shows much lower matched fraction of $\sim$ 60$\%$ but even higher significance $\sim$ 25. A loosening of the constraints in the criterion of redshift difference introduces higher random matched fraction, and thus higher false detection rates. It thus follows that the photo-$z$ accuracy has a significant impact on the group finding. Moreover, we find that the low matched fraction $\sim$ 60$\%$ from the zCOSMOS catalog mainly results from the problems of its survey incompleteness, in some areas at lower redshift that accounts for the unmatched fraction of $\sim$ 13$\%$, and for the flux-limited, high redshift groups that accounts for $\sim$ 29$\%$. Furthermore, both PFOF COSMOS and PS1MD04 catalogs also have matched fractions close to their respective maximum matched fractions, 89.9 and 91.4$\%$, suggesting that our subset optimization is successful.

\begin{figure*}
 \begin{center}

 \includegraphics[angle=0,scale=0.9]{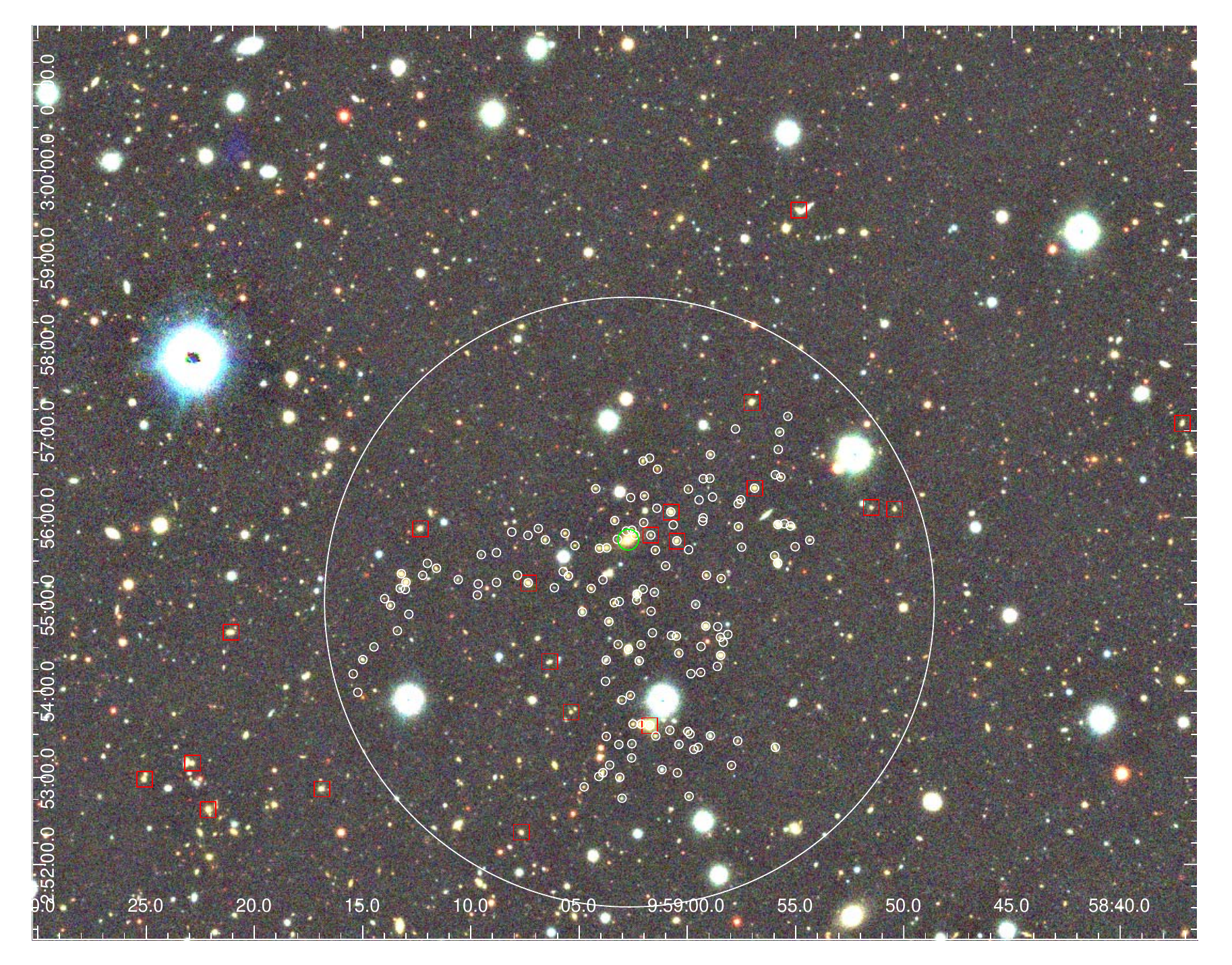}
  \caption{RXJ0959.0+0255: A PS1 image shows a PFOF detection ($z$ = 0.425) in PS1MD04 on top of an X-ray cluster position ($z$ = 0.3494). Members are marked in white while the BCG is indicated by the green circle. In addition, the larger white circle corresponds to 1 $\Mpc$ based on the cluster redshift, and spectral-$z$ galaxies are denoted by red squares in $z$ range between 0.345 and 0.355. The map size is $\sim$ 13.4' $\times$ 10.5'.
   }\label{fig:RXJ}
 \end{center}
\end{figure*}

\begin{figure*}
 \begin{center}

 \includegraphics[angle=0,scale=0.9]{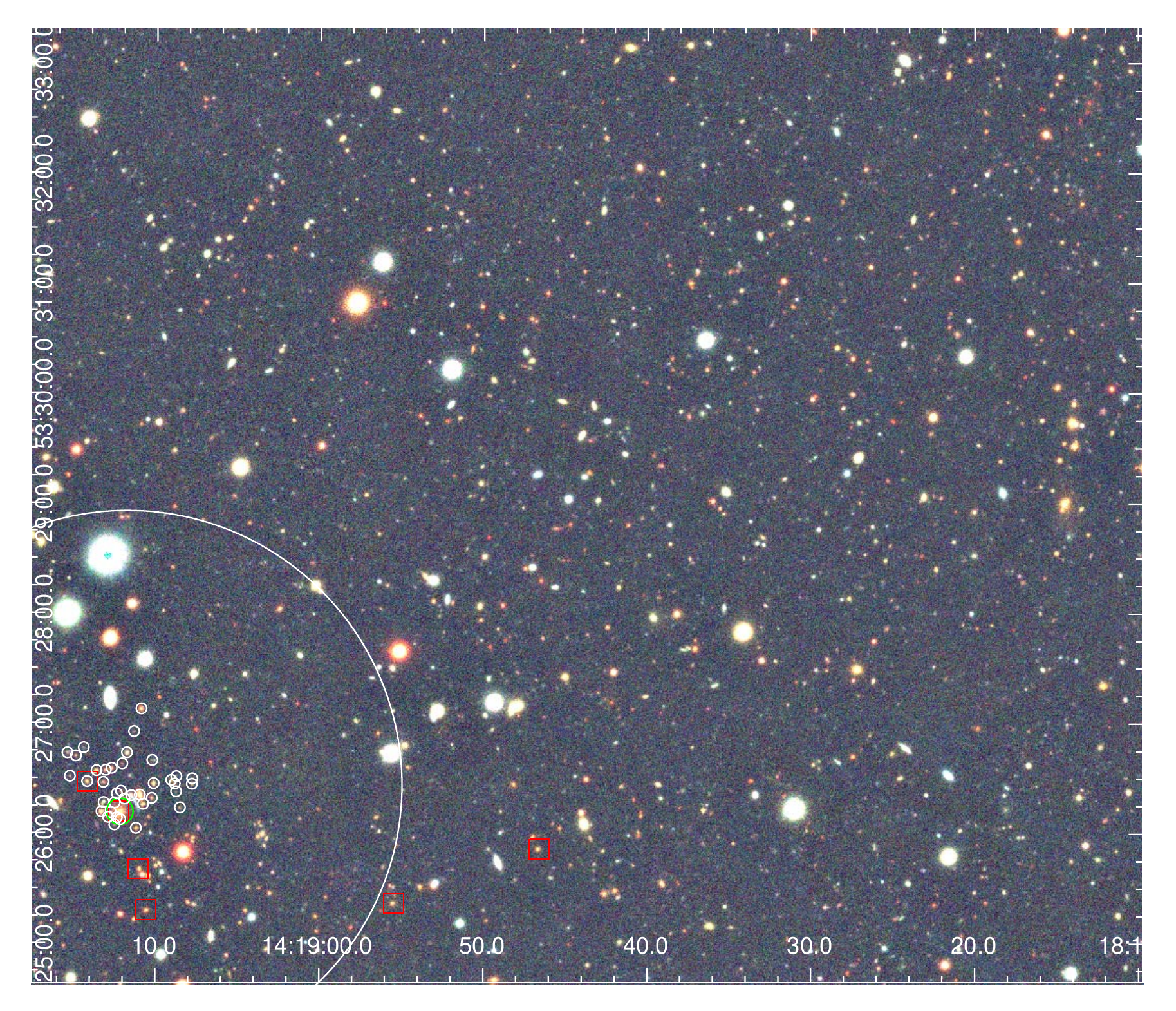}
  \caption{eMACSJ1419.2+5326: Similar to Figure~\ref{fig:RXJ}, but a PFOF detection ($z$ = 0.687) in PS1MD07 on top of an X-ray cluster position ($z$ = 0.6384), and spectral-$z$ galaxies (red squares) are in $z$ range between 0.631 and 0.645. The X-ray cluster is on the corner of skycell 80. The map size is $\sim$ 16.9' $\times$ 8.6'.
   }\label{fig:eMACS}
 \end{center}
\end{figure*}

We also plot the location, angular size, and redshift of X-ray clusters (open circles) and zCOSMOS groups (solid circles) from zCOSMOS catalog \cite{kno12} in (a), from PFOF COSMOS catalog in (b), and from PFOF PS1MD04 catalog in (c), respectively, in Figure~\ref{fig:XrayMatchMap} to demonstrate the detail of matched results. The angular size for the X-ray clusters is $R_{200}$ and is the maximum member-group center distance for PFOF groups. From the map in (a), it is not surprising that the matched zCOSMOS groups have comparable size to X-ray clusters. Comparing results between the two different photo-$z$ samples, i.e. (b) and (c), groups in PS1MD04 have more fragmentation. This implies that to acquire a higher matched fraction, PFOF groups tend to be more fragmented. This is because overmergers can often yield group positions departing from their real locations but also include too many false members to have group redshifts deviating from their real redshifts.

\cite{ebe13} recently presented the results of a pilot study for the extended MACS survey (eMACS), Which aims to expand the MACS cluster survey to higher redshift and lower X-ray fluxes by combining the two large-area imaging datasets introduced in the preceding
sections: the ROSAT All-Sky Survey (RASS), including Bright and Faint Source Catalogs (BSC and FSC), and the PS1 3$\pi$ survey. They apply no additional constraints regarding X-ray flux, spectral hardness ratio, or photon statistics and lower the redshift threshold to $z$ $>$ 0.3 to extend the probed luminosity range to poorer systems. Examination of PS1/MDS images for 41 BSC and 200 FSC sources combined with dedicated spectroscopic follow-up observations results in a sample of 11 clusters with estimated or spectroscopic redshifts of $z$ $>$ 0.3. Among those clusters, RXJ0959.0+0255 (gri, $z$ = 0.3494) is in PS1MD04, and  eMACSJ1419.2+5326 (riz, $z$ = 0.6384) in PS1MD07, and we thus use them to check PFOF identifications in these two fields. Figure~\ref{fig:RXJ} shows the rgb PS1 image (z-Band for red, r-Band for green, and g-Band for blue) around RXJ0959.0+0255 and the PFOF detected group which has a mean photo-$z$ redshift $z$ $\sim$ 0.43 in PS1MD04. The white circles denote PFOF group galaxies, the blue circle indicates the BCG of the group, the large white circle corresponds to a radius of 1 $\Mpc$, and red squares are spectral-$z$ galaxies. Similar to Figure~\ref{fig:RXJ}, Figure~\ref{fig:eMACS} shows the image around eMACSJ1419.2+5326, the PFOF detected group which has a mean photo-$z$ redshift $z$ $\sim$ 0.69 in PS1MD07. In these two cases, both clusters are detected by PFOF. It is demonstrated again that PFOF is capable of finding clusters even for surveys with relatively ``poor'' photo-$z$ accuracy, such as PS1 survey.

\subsection{Estimation on the Recovering and Contamination Rate for Red and Blue Group Galaxies of PFOF}

It is well-known that the accuracy of photo-$z$ is correlated with colors of galaxies, and therefore now we turn to discuss how well PFOF can recover blue and red group galaxies respectively, by using the EGS described in Section~\ref{EGS} and the PS1MD07 group catalog in Section~\ref{PS1MD07}. We select PFOF group galaxies with spectral-$z$ in groups with richness $N$ $\geq$ 4 and redshift between 0.5 and 1.0, and also identify their corresponding spectral-$z$ groups and group galaxies in these corresponding spectral-$z$ groups to obtain the recovery rate as a function of color. We find that in terms of rest-frame $U-B$ color or $(U-B)_0$, where galaxies with $(U-B)_0$ $\geq$ 0.9 are red and others are blue, in the EGS group catalog, PFOF recovers 43 red and 99 blue spectral-$z$ group galaxies from the original 63 red and 147 blue DEEP2 EGS group galaxies, and the recovery rate for red and blue galaxies in groups are 68.3$\%$ and 67.3$\%$, respectively. Similarly, in the PS1MD07 group catalog, we find that the recovering rate for red galaxies is 73.6$\%$ (89/121) and for blue ones is 71.5$\%$ (264/396). In other words, PFOF detection is not biased to blue or red galaxies. This demonstrates that PFOF does not have a preferential color selection to identify a galaxy as a group member. The ratio of red to blue group galaxies is $\sim$ 0.27 in EGS and $\sim$ 0.28 in PS1MD07. We also compute the contamination rate in terms of color for PFOF group galaxies with spectral-$z$ in groups with richness $N$ $\geq$ 4 and redshift between 0.5 and 1.0. In the EGS group catalog, we find that for red galaxies, 32 out of 75 ($\sim$ 42.7$\%$) galaxies in PFOF group galaxies with spectral-$z$ are actually field galaxies according to the DEEP2 group catalog, and for blue, 143 out of 242 ($\sim$ 59.1$\%$) are field galaxies. On the other hand, in the PS1MD07 group catalog, the contamination rate for red galaxies is $\sim$ 55.5$\%$ (111/200) and for blue ones is $\sim$ 66.8$\%$ (532/796). That the contamination rate is less for red galaxies is expected due to better photo-$z$ for red than for blue galaxies in general.

The spectral-z samples used in this paper as the training sets are mainly from the DEEP2 survey with a magnitude cut of R $<$ 24.1 and the zCOSMOS sample with i $<$ 22.5. The two samples may be biased toward emission-line galaxies in particular in the faint end regime where the red galaxies do not have enough S/N in the absorption features. On the other hand, due to the poorer photo-$z$ accuracy for blue galaxies compared to red ones, a training sample that contains more blue galaxies may be helpful for calibrating the group finder parameters. There is probably no perfect solution for a training sample to get rid of such bias. PFOF is inevitably also affected by the training sample bias. However, depending on the science goals, one can use different samples for the training purpose when applying PFOF. Optimizing the parameters to recover the blue group members is one of the strengths of PFOF, unlike many other group-finding methods that are biased toward red galaxies. Ideally, we may take into account the selection function of individual galaxies in the spectroscopic sample by giving the weighting when doing the optimization. The main purpose of this paper is to demonstrate the performance of PFOF, and we thus assume no selection effect in the sub-samples.

\section{Summary}
We extend analyses of the PFOF group finder from the previous paper of \cite{liu08}. We start by briefly reviewing the PFOF algorithm and then illustrating definitions of purity and completeness in PFOF. We adopt an optimization measure $\widetilde{g_1}$ introduced by \cite{kno12}. With the Durham mock catalog for Pan-STARRS Medium Deep Survey, we demonstrate PFOF performance for various photo-$z$ accuracies. In addition, to reduce any dependency of our group catalogs on the details of the mock catalogs, we adopt a calibration method, called subset optimization, by using spectroscopically identified groups from observational data as a training set to optimize PFOF grouping. The method is examined by using the mock with a similar sampling rate and is shown to be feasible. We then apply subset optimization to observational datasets, considering photo-$z$ accuracies ranging from ``good'' $\sigma_{\Delta z/(1+z_s)}$ $\sim$ 0.01 in COSMOS, ``medium'' $\sim$ 0.03 in EGS, to relatively ``poor'' $\sim$ 0.06 in PS1MD04 and PS1MD07, to illustrate the performance of the PFOF in terms of photo-$z$ accuracy. Moreover, we also match PFOF groups to NEWTON-XMM X-ray clusters to have an alternative performance check. In the end, we estimate the recovering rate for red and blue group galaxies to demonstrate that PFOF is not biased by color. Our results are summarized as follows.

(1) To assess how PFOF performance varies with photo-$z$ accuracy $\sigma_{\Delta z/(1+z_s)}$, we make use of the Durham mock catalog with simulated photo-$z$ to study optimal purity and completeness as a function of $\sigma_{\Delta z/(1+z_s)}$ as shown in Figure~\ref{fig:pcSig}. We find that purity or completeness drops by only $\sim$ 20$\%$ when $\sigma_{\Delta z/(1+z_s)}$ deteriorates from 0.03 to 0.07, to the expected range of redshift accuracy for PS1 photo-$z$.

(2) Using a simulated $\sigma_{\Delta z/(1+z_s)}$ $\sim$ 0.06 with a selection cut, where i $\leq$ 24.1, $z$ $\leq$ 1.4, and richness $N$ $\leq$ 4, as an example, we illustrate the performance of the PFOF in detail. For this case, the optimal values of $p_1$ and $c_1$ are 54$\%$ and 49$\%$, respectively. We plot spatial distribution of PFOF groups over-plotted with mock groups for halo mass $M_h$ $>$ $10^{14} \Msun$ in Figure~\ref{fig:7SqrDegMap}. It is found that PFOF detects 112 out of 116 mock clusters with roughly correct redshifts, where the successful detection is to have at least 5 member galaxies from the mock groups. The 4 undetected clusters are at high redshift ($z$ $>$ 1) and are low richness groups (N $\leq$ 5). In addition, it is found that clusters tend to be fragmented by PFOF, i.e.  116 mock clusters fragmented into 875 PFOF groups, leading to a condition with high purity but low completeness.

(3) When purity $p_1$ and completeness $c_1$ are expressed in terms of redshift $z$ and halo mass $M_h$ in Figure~\ref{fig:p1c1}, we find that in the high mass range, PFOF groups are pure, but their corresponding mock groups are not complete, consistent with results described in (2). On the other hand, at lower $M_h$ and higher $z$, we have the opposite trend that mock groups are more complete but their corresponding PFOF groups are less pure. The high purity and low completeness in the high halo mass region seem to disagree with the optimal values of $p_1$ and $c_1$ mainly because the abundant low N groups dominate the global measures and compromise the high-mass-end results.

(4) In addition to the definition we adopt for purity $p_1$ and completeness $c_1$ \citep{ger05,ger12,kno09,kno12}, we also use the definition adopted in ORCA \citep{mur12} for purity, defined as the fraction of galaxies assigned to the cluster that are members of the host halo, and for a complete group, defined as that a halo is detected if at least $N_{min}$ galaxies are identified, even if they are shared between multiple ORCA clusters, to demonstrate that for the same group catalog, performance based on different definitions can vary significantly.

(5) From the mock test, when we increase the sample depth from 24.1 to 25.8 for PFOF grouping, the performance does not decrease significantly, only by $\sim$ 8$\%$. When we adopt an absolute magnitude cut instead of a limiting magnitude cut for the same number density, we can gain the performance in $\sim$ 20$\%$.

(6) We develop an optimization method, called the subset optimization, for PFOF, and make use of the mock catalog to simulate two different cases to test the idea and demonstrate from both cases that the optimal linking lengths and threshold obtained from the sub-sample coincide with those obtained from the full-sample. In other words, the subset optimization is a feasible methodology.

(7) By using the 10k zCOSMOS groups as the training set, the optimal purity $p_1$ and completeness $c_1$ from the sub-sample are 66$\%$ and 70$\%$ for COSMOS galaxies with ``good'' photo-$z$ accuracy $\sim$ 0.01, and are 40$\%$ and 57$\%$ for PS1MD04 with poorer photo-$z$ accuracy $\sim$ 0.05.

(8) By using the DEEP2 EGS groups to train the linking lengths, we obtain optimal values of $p_1$ = 55$\%$ and $c_1$ = 63$\%$ for the EGS catalog with ``medium'' photo-$z$ accuracy $\sim$ 0.03, and optimal values of $p_1$ = 53$\%$ and $c_1$ = 47$\%$ for PS1MD07 catalog with ``poorer'' photo-$z$ accuracy $\sim$ 0.054. The optimal performance from these two group catalogs is consistent with that derived from Durham mock catalog with similar simulated photo-$z$ accuracies.

(9) To further examine the performance of the PFOF, we match two PFOF group catalogs, COSMOS and PS1MD04, to the XMM-NEWTON plus Chandra X-ray catalog to understand how well we may recover X-ray sources. The matched fraction for PFOF COSMOS groups is $\sim$ 85$\%$ with significance 5.0, and the matched fraction for PFOF PS1MD04 groups is comparable $\sim$ 87$\%$ but with lower significance 1.5, where the significance is defined as the ratio of the matched fraction from the X-ray catalog to the average fraction from the random catalogs. The matched fractions 85 and 87$\%$ are close to their respectively maximum matched fractions 89.9 and 91.4$\%$, implying that the subset optimization is successful.

(10)To assess how well PFOF can recover blue and red group galaxies, we make use of group galaxies with spectroscopic redshift in the EGS and PS1MD07 group catalogs to estimate the recovering rate for red and blue group galaxies. We find that the recovering rate is roughly the same for blue and red group galaxies $\sim$ 70$\%$ in both the EGS and PS1MD07 samples, demonstrating that PFOF detection has no color bias for galaxy groups.

To conclude, we find that the performance of PFOF does not drop significantly with the photo-$z$ accuracy in the range between 0.03 and 0.07 for the mock tests. We also find that the subset optimization is successful for PFOF group finding, and PFOF can be applied to a real sample with PS1-like photo-$z$ accuracy $\sigma_{\Delta z/(1+z_s)}$ $\sim$ 0.05 with the optimal purity and completeness reaching $\sim$ 0.5 from the observational datasets. In addition, we also show that PFOF can detect blue galaxies well and the recovery rate is roughly the same for red ($\sim$ 68.3$\%$) and for blue ($\sim$ 67.3$\%$) group galaxies with spectral-$z$ in the EGS and is $\sim$ 73.6$\%$ for red and $\sim$ 71.5$\%$ in PS1MD07, and thus, demonstrate the capability of PFOF to find blue members in a group or cluster.

The purpose for this paper is to illustrate the capability of PFOF to find groups and clusters with photo-$z$ accuracy achieved by PS1-like surveys, but not to release catalogs at this point. In the near future, we plan to release group catalogs of PS1 MDS by combining grouping results from ORCA \citep{mur12} and the PFOF algorithms for future science studies. In addition, to make PFOF group finding applicable, we find that a subset with high sampling rate of spectral-$z$ catalog is necessary, and we thus strongly suggest that a photometric survey should be accompanied with a spectral-$z$ survey with a high sampling rate.

~

\emph{Acknowledgements}-

We thank R. Bower, M. Takada, and M. Oguri for helpful discussions on our algorithm and PFOF applications, and P. Price for the valuable comments. We also thank Brian F. Gerke for providing us the DEEP2 group catalogs for the PFOF training. The work is supported in parts by the National Science Council of Taiwan under the grant NSC101-2811-M-002-075, NSC99-2112-M-001-003-MY3, NSC101-2112-M-001-011-MY2, and NSC101-2628-M-008-002-.

The Pan-STARRS1 Surveys (PS1) have been made possible through contributions of the Institute for Astronomy, the University of Hawaii, the Pan-STARRS Project Office, the Max-Planck Society and its participating institutes, the Max Planck Institute for Astronomy, Heidelberg and the Max Planck Institute for Extraterrestrial Physics, Garching, The Johns Hopkins University, Durham University, the University of Edinburgh, Queen's University Belfast, the Harvard-Smithsonian Center for Astrophysics, the Las Cumbres Observatory Global Telescope Network Incorporated, the National Central University of Taiwan, the Space Telescope Science Institute, the National Aeronautics and Space Administration under Grant No. NNX08AR22G issued through the Planetary Science Division of the NASA Science Mission Directorate, the National Science Foundation under Grant No. AST-1238877, and the University of Maryland.


\end{document}